\documentclass{aa}     

\usepackage{graphicx}
\usepackage{txfonts}
\usepackage{natbib}
\bibpunct{(}{)}{;}{a}{}{,}

\newcommand{\Mpc}{$h^{-1}$\thinspace Mpc}

\newcommand{\vmh}{h^{-1}\mathrm{Mpc} }

\def\apjl{ApJL}
\def\apjs{ApJS}

\begin{document}

\title{SDSS DR7 superclusters} 
\subtitle{Morphology} 
\author { M.  Einasto\inst{1} \and L.J. Liivam\"agi\inst{1,2} \and E.  Tago\inst{1} 
\and E.  Saar\inst{1}  \and  E.  Tempel\inst{1,2}  \and J.  Einasto\inst{1}
\and V.J. Mart\'{\i}nez\inst{3} \and P. Hein\"am\"aki\inst{4}
}
\institute{Tartu Observatory, 61602 T\~oravere, Estonia
\and 
Institute of Physics, Tartu University, T\"ahe 4, 51010 Tartu, Estonia
\and 
Observatori Astron\`omic, Universitat de Val\`encia, Apartat
de Correus 22085, E-46071 Val\`encia, Spain
\and 
Tuorla Observatory, University of Turku, V\"ais\"al\"antie 20, Piikki\"o, Finland
}
\authorrunning{M. Einasto et al. }
\offprints{M. Einasto}
\date{ Received    / Accepted...   }
\titlerunning{Morphology of superclusters}

\abstract{}
{We study the morphology of a set of superclusters drawn from the SDSS DR7. }
{ 
We calculate the luminosity density field to determine superclusters from a 
flux-limited sample of galaxies from SDSS DR7, and select superclusters with 300 and 
more galaxies for our study. We characterise the morphology of superclusters 
using the fourth Minkowski functional $V_3$, the morphological signature (the curve 
in the shapefinder's $K_1$-$K_2$ plane) and the shape parameter (the ratio of the
shapefinders $K_1/K_2$).  We investigate the supercluster sample using 
multidimensional normal mixture modelling. 
We use Abell clusters to identify our 
superclusters with known superclusters and to study the 
large-scale distribution of superclusters.
}
{
The superclusters in our sample form three chains of 
superclusters; one of them is the Sloan Great Wall. Most superclusters have 
filament-like overall shapes. Superclusters can be divided into two sets; more 
elongated superclusters are more luminous, richer, have larger diameters, and 
a more complex fine structure than less elongated superclusters. The fine 
structure of superclusters can be divided into four main morphological types: 
spiders, multispiders, filaments, and multibranching filaments.
We present the 2D and 3D distribution of galaxies and rich groups,  
the fourth Minkowski functional, and the morphological signature for all 
superclusters.  
}
{Widely different morphologies of superclusters show that their evolution has 
been dissimilar. 
A study of  a larger sample of 
superclusters from observations and simulations is needed 
to understand the morphological variety of superclusters
and the possible connection between the morphology of superclusters
and their large-scale environment.
}

\keywords{cosmology: observations -- cosmology: large-scale structure
of the Universe -- galaxies: clusters: general}

\maketitle

\section{Introduction} \label{sect:intro} 

The most remarkable feature of the megaparsec-scale matter distribution in the 
Universe is the presence of the cosmic web -- the network of galaxies, groups, and 
clusters, connected by filaments 
\citep{1978MNRAS.185..357J,1978ApJ...222..784G,zes82,1986ApJ...302L...1D}. The 
formation of a web of galaxies and systems of galaxies is predicted in any 
physically motivated theory of the formation of structure
 in the Universe 
\citep[see, e.g.,][]{1996Natur.380..603B}. In 
this scenario
galaxies and galaxy systems form because of initial density perturbations on 
different scales. Perturbations on a scale of about 100~\Mpc\ 
($H_0=100 h \mathrm{km\,s^{-1}Mpc^{-1}}$) 
     give rise to the 
largest systems of galaxies -- rich superclusters. At larger scales 
dynamical evolution proceeds at a slower rate and  superclusters 
have retained the memory of the initial conditions of their formation and of the 
early evolution of structure 
\citep{1980MNRAS.193..353E, zes82, 1987Natur.326...48K}.

Numerical simulations show that high-density peaks in the density
distribution (the seeds of supercluster cores) are seen already at
very early stages of the formation and evolution of structure
\citep{2010AIPC.1205...72E}.  These are the locations of the formation
of the first objects in the Universe
\citep[e.g.][]{2004A&A...424L..17V,2005ApJ...635..832M,
  2005ApJ...620L...1O}. Observations have already found
superclusters at high redshifts
\citep{2005MNRAS.357.1357N,2007MNRAS.379.1343S,2008ApJ...684..933G,tanaka09}.
Recently, the XMM-Newton
satellite
 follow-up for the validation of Planck cluster candidates 
led to the discovery of two massive, previously unknown superclusters of 
galaxies \citep{2011arXiv1101.2024P}. These are likely the first superclusters 
discovered through the Sunyaev-Zeldovich  effect. 

Superclusters are important tracers of dark and baryonic matter in
the Universe 
\citep{2005A&A...434..801Z,2005MNRAS.363...79G,
  2008MNRAS.385.1431H,2009ApJ...695.1351B,2009MNRAS.396...53P,
  2011arXiv1102.4617S}.
Studies of superclusters and of the supercluster-void network
have demonstrated the presence of a characteristic scale in the
distribution of rich superclusters
\citep{1994MNRAS.269..301E,1997Natur.385..139E}. This was probably an
early hint of baryon acoustic oscillations \citep{2010MNRAS.401.2477H}.

To search for superclusters and to understand their properties, we need to know 
how to identify
them and how to quantify their properties 
\citep{2010MNRAS.409..156B}. 
Several 
 methods have been proposed to study the 
cosmic web \citep[][and references 
therein]{2004ApJ...606...25B,2010A&A...510A..38S,2010MNRAS.tmp.1270A, 
2011MNRAS.tmp..511S, 2011MNRAS.tmp..530S}. One approach is to determine cosmic 
structures (in our study -- superclusters of galaxies) using the density field and 
to study their morphology with Minkowski functionals and shapefinders 
\citep[][and references therein] 
{1997ApJ...482L...1S,1998ApJ...508..551S,bas03,2003MNRAS.343...22S,sss04,e07}.
\citet{1983ARA&A..21..373O} gave a review of the early studies of superclusters. 
Supercluster catalogues  compiled based  on data about clusters of 
galaxies were published by 
\citet{1993ApJ...407..470Z, 1994MNRAS.269..301E, 1995A&AS..113..451K, 
e2001}.  
Recent deep surveys of galaxies  
\citep[as 2dFGRS and SDSS, see][]{col03,abaz08} introduced a new era in the studies of 
the large-scale structure of the Universe, where systems of galaxies can be 
studied in unprecedended detail. A number of 
supercluster catalogues have been compiled with these data, 
\citep[][and references 
therein]{bas03,2003A&A...405..425E, 2004MNRAS.352..939E,
2006A&A...459L...1E, 2007A&A...462..811E,
2010arXiv1012.1989J, 2011arXiv1101.1961L}.

The overall morphology of superclusters has been studied by several authors 
\citep{kbp02, bas03, 2011MNRAS.411.1716C}, we refer to \citet{2007A&A...462..397E} 
for a review of the properties of superclusters. The shapes and sizes of 
superclusters can be used to compare the observed 
superclusters with those obtained from cosmological simulations 
\citep{kbp02,2007A&A...462..397E}. \citet{nichol06} showed that the higher order 
correlation functions 
of the 2dFGRS do not agree with 
those found in numerical simulations 
\citep[but this fact can be explained by non-Gaussian initial 
density fields,][]{1996ApJ...462L...1G}. This discrepancy may be caused by the 
unusual morphology of  
one of the richest superclusters in the 2dFGRS, the supercluster SCl~126 
 \citep{e07, e08} from the catalogue of superclusters
 by \citet[][hereafter E01]{e2001}.
The morphology of superclusters may be used to distinguish between
different cosmological models \citep{kbp02}.

Superclusters contain structures with a wide range of densities, from high-density 
cores of rich clusters to low-density filaments between clusters and groups. 
This makes them  ideal laboratories to study processes that affect the 
evolution of galaxies,  groups, and clusters of galaxies. A number of studies 
have already  shown that the supercluster environment affects the properties of 
galaxies, groups, and clusters located there \citep{e2003b,2004ogci.conf...19P, 
2005A&A...443..435W, 2006MNRAS.371...55H, e07b, 
2008MNRAS.388.1152P, tempel09, 2010ASPC..423...81F, 2011A&A...529A..53T,
2011arXiv1105.1632E}. Other 
evidence about the influence of the large-scale environment of galaxies on their 
properties comes from the study of the properties of galaxies in void walls 
\citep{2008MNRAS.390L...9C} and from the study of the large-scale environment 
of quasars \citep{Lietzen2009}.  
A detailed information on the morphology 
of superclusters is needed
to find out whether that may also be an 
important environmental factor in shaping the properties of galaxies and groups 
of galaxies in superclusters \citep[see also][]{2010MNRAS.tmp.1270A}.

\citet{e07} showed
 that the morphology of a typical poor
supercluster can be described as a ``spider'' -- a system of several 
filaments growing from one
concentration centre (a rich cluster). The Local Supercluster is an
example of a typical ``spider'. 
Rich superclusters can be described as ``multispiders'',
where several high-density clumps are connected by lower-density
filaments. One very rich supercluster (SCl~126) was described as a
multibranching filament that consists of a rich filament of a quite
uniform high density with poorer filaments for branches. 
This work concerned only a few of the richest supercluster
while the present analysis enables us to test
 the classification of the morphology of superclusters.

The goal of the present paper is to study in detail the morphology of a large 
sample of superclusters drawn from the 7th 
data release of the Sloan Digital Sky Survey (SDSS). 
The superclusters are determined using
the global luminosity field and their  morphologies are quantified with
 the Minkowski functionals and shapefinders. 
The superclusters are divided  into two sets using 
multidimensional normal mixture modelling, applying the
{\it Mclust} package for clustering
and classification \citep{fraley2006} from {\it R}, an 
open-source free statistical environment developed under the GNU GPL 
\citep[][\texttt{http://www.r- project.org}]{ig96}.  
The cluster membership of superclusters and their large-scale distribution is 
analysed, and 
a short description of individual superclusters is given. 
The data are described in Sect.~\ref{sect:data}, 
the methods in Sect.~\ref{sect:method}, and the results
in Sect.~\ref{sect:results}. In Sect.~\ref{sect:discussion} 
the selection effects in our sample are discussed,
and a comparison with other studies is given.
The study is summarised  in Sect.~\ref{sect:conc}.
Interactive 3D 
models of the richest superclusters in our sample 
can be found on our web page: 
\texttt{http://www.aai.ee/$\sim$maret/SDSSsclmorph.html}.

We assume  the standard cosmological parameters: the Hubble parameter $H_0=100~ 
h$ km s$^{-1}$ Mpc$^{-1}$, the matter density $\Omega_{\rm m} = 0.27$, and the 
dark energy density $\Omega_{\Lambda} = 0.73$.

\section{Data}
\label{sect:data}

We selected the MAIN galaxy sample of the 7th data release of the Sloan
Digital Sky Survey \citep{ade08,abaz08}
with the apparent $r$ magnitudes $12.5 \leq r \leq 17.77$,   excluding
duplicate entries. The sample  is described in detail in
\citet{2010A&A...514A.102T}, hereafter T10. 
We corrected the redshifts
of galaxies for the motion relative to the CMB and computed the co-moving
distances \citep{mar03} of galaxies.

The absolute magnitudes of galaxies are determined in the $r$-band $M_r$ with
$k$-correction for the SDSS galaxies calculated with the KCORRECT
algorithm \citep{blanton03a,2007AJ....133..734B}. In addition, we
applied the evolution corrections, using the luminosity evolution model of
\citet{blanton03b}. The magnitudes correspond to the rest-frame at the
redshift $z=0$.

The first step is to determine groups and clusters of galaxies with the friends-
of-friends algorithm, where a galaxy belongs to a group of galaxies  if this 
galaxy has at least one group member galaxy closer than the selected linking 
length. The linking length along with the distance was increased, to take into 
account selection effects, when constructing a group catalogue for a flux-
limited sample. As a result, the maximum sizes and velocity dispersions of 
groups are similar at all distances. For details and for the group catalogue we 
refer the reader to T10.
\footnote{
The T10 group catalogue is
available in electronic form at the CDS via anonymous ftp to
cdsarc.u-strasbg.fr (130.79.128.5) or via
\texttt{http://cdsweb.u-strasbg.fr/cgi-bin/qcat?J/A+A/ 514/A102}.
}

To determine the luminosities of groups and to calculate the luminosity density 
field we have also to correct for the luminosities of galaxies that lie outside 
of the survey magnitude range.
The calculation of luminosities is described 
in Appendix~\ref{sec:DF} \citep[details of this calculation are given also 
in][]{2011A&A...529A..53T}.

\begin{figure}[ht]
\centering
\resizebox{0.45\textwidth}{!}{\includegraphics*{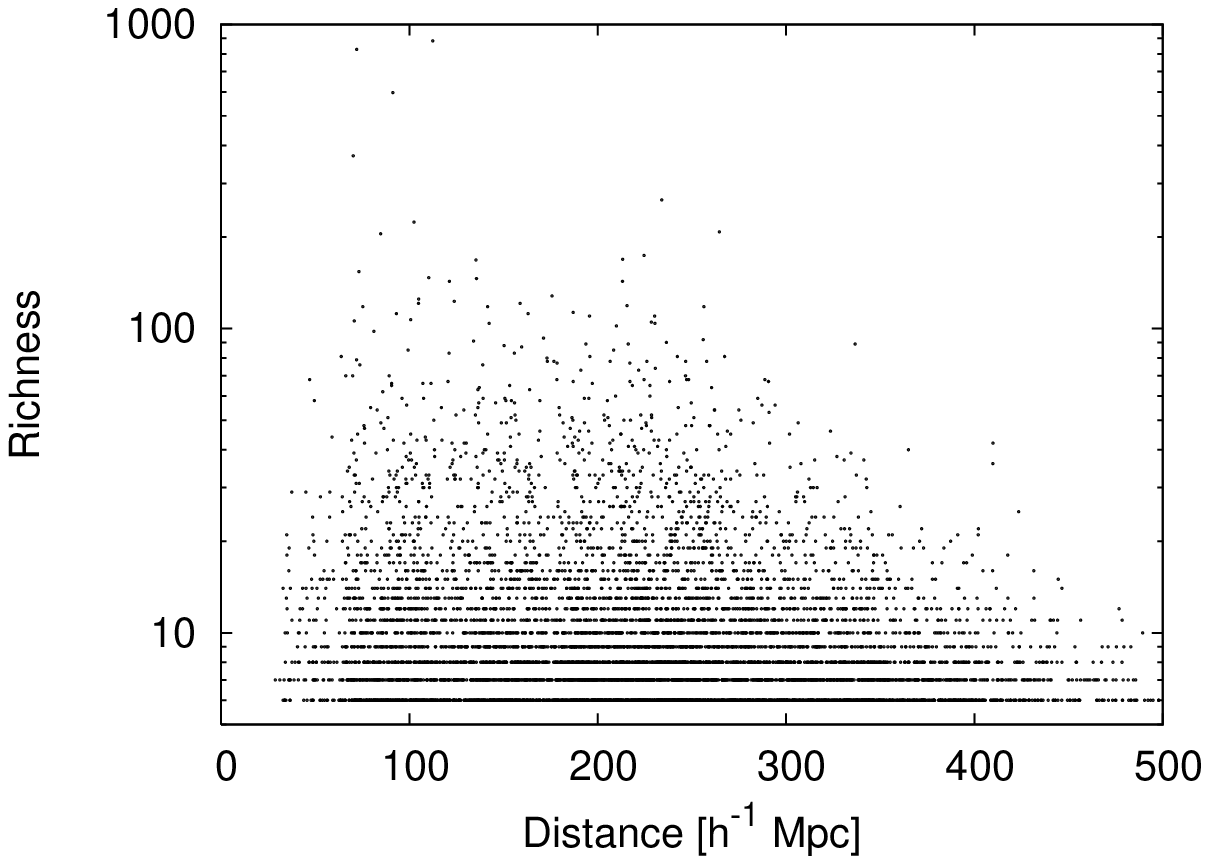}}
\caption{Richness of groups vs. their distance in the T10 group catalogue.
Only the data for groups with at least six member galaxies are
plotted. This plot shows that at distances larger than $\approx$ 320~\Mpc\
almost no  rich group  is visible.   
}
\label{fig:rich6}
\end{figure}

In the final flux-limited group catalogue the richness of groups decreases 
rapidly at distances $D > 320$~\Mpc\ because of selection effects. This effect 
is seen in Fig.~\ref{fig:rich6}. At small distances, $D < 70$~\Mpc\, luminosity 
weights are large owing to the absence of very bright galaxies.  Therefore  we 
chose for the present analysisa subsample of galaxies and galaxy systems in the 
distance interval 90~\Mpc\ $\le D \le $ 320~\Mpc\, where the selection effects 
are small. 

\begin{table*}[ht]
\caption{Data of the superclusters.}
\label{tab:scldata}  
\begin{tabular}{rrrrrrrrrrrrrr} 
\hline\hline 
(1)&(2)&(3)&\multicolumn{1}{c}{(4)}&(5)& (6)&(7)&(8)& (9)& (10)&(11)& (12)& (13)&(14)\\      
\hline 
 $SCl_{ID}$ &$ID_{E01}$&\multicolumn{1}{c}{ID}& $N_{\mbox{gal}}$ &$N_{\mbox{30}}$ & $d_{\mbox{peak}}$  & $L_{\mbox{tot}}$ 
& $Diam.$& $R_{\mbox{magV}}$& $N_{\mbox{galV}}$  & $V_{3,\mathrm{max}}$ & $K_1$ & $K_2$ & $K_1$/$K_2$ \\
   &          &            &       &    & Mpc/$h$&$10^{10}h^{-2} L_{\sun}$& Mpc/$h$ &mag&  &   & & & \\
\hline
   1& 162     & 239+027+009 &1038  &  4 & 264.5  &  1591.5&  50.3 &  -19.70 & 916   &  2 & 0.077 & 0.163 & 0.48   \\ 
  10& 160    & 239+016+003  &1463  &  2 & 111.9 &    680.2 & 22.8  &  -17.50  &1455  &  1 & 0.026 & 0.024 & -      \\
  11& 154     & 227+006+007 &1222  &  5 & 235.0  &  1476.0 & 35.4 &  -19.30 & 1180  &  3 & 0.051 & 0.056 & 0.92   \\ 
  24& 111     & 184+003+007 &1469  &  4 & 230.4  &  1768.2&  56.4 &  -19.25 & 1419  &  6 & 0.077 & 0.163 & 0.48   \\
  55& 111     & 173+014+008 &1306  &  4 & 242.0  &  1773.0&  50.3 &  -19.55 & 1155  &  5 & 0.084 & 0.198 & 0.43  \\
  60& 160    & 247+040+002  &1335  &  3 &  92.0 &    527.4 & 21.2  &  -17.30  &1275  &  1 & 0.010 & 0.020 & -      \\
  61& 126     & 202-001+008 &3056  & 15 & 255.6  &  4315.3& 107.0 &  -19.25 & 3048  & 13 & 0.130 & 0.460 & 0.28   \\
  94& 158     & 230+027+006 &1830  & 10 & 215.4  &  2263.4&  54.6 &  -19.40 & 1632  &  9 & 0.110 & 0.399 & 0.28  \\
 336& 109     & 172+054+007 &1005  &  6 & 170.0  &  1003.6 & 53.3 &  -19.30 & 832   &  4 & 0.081 & 0.250 & 0.33   \\
 350& 160     & 230+008+003 & 955  &  5 & 105.8  &   436.3 & 23.0 &  -17.50 & 947   &  2 & 0.029 & 0.045 & 0.65   \\
  38&  95     & 167+040+007 &586   &  5 & 224.4 &    660.7 & 22.5 &   -19.40 & 507  &  2 & 0.028 & 0.033 & 0.84   \\
  64& 164     & 250+027+010 &619   &  2 & 301.7 &   1305.4 & 55.7 &   -19.70 & 618  &  3 & 0.087 & 0.230 & 0.38   \\
  87&  -      & 215+048+007  & 445  &  1 & 213.4 &    477.8 & 21.6  &  -19.00  & 445  &  2 & 0.036 & 0.033 & 1.09   \\
 136& 271     & 189+017+007 &504   &  5 & 212.1 &    523.2 & 20.7 &   -19.25 & 446  &  2 & 0.033 & 0.017 & 0.97   \\
 152& 160    & 230+005+010  & 423  &  2 & 301.6 &    907.5 & 32.8  &  -19.75  & 422  &  3 & 0.068 & 0.081 & 0.84   \\
 189&  -      & 126+017+009  & 433  &  1 & 267.2 &    771.0 & 43.7  &  -19.70  & 409  &  4 & 0.066 & 0.195 & 0.35   \\
 198&  82    & 152-000+009  & 473  &  1 & 284.7 &    863.9 & 38.7  &  -19.80  & 454  &  4 & 0.056 & 0.086 & 0.65   \\
 223& 111    & 187+008+008  & 462  &  2 & 268.3 &    703.7 & 34.0  &  -19.60  & 442  &  3 & 0.054 & 0.139 & 0.39   \\
 228& 133     & 203+059+007 &643   &  4 & 210.6 &    644.0 & 31.1 &   -19.10 & 612  &  2 & 0.043 & 0.050 & 0.86   \\
 317&  -      & 156+010+010  & 351  &  1 & 321.6 &    846.7 & 36.0  &  -20.00  & 345  &  5 & 0.077 & 0.226 & 0.34   \\
 332& 106    & 175+005+009  & 333  &  1 & 291.0 &    664.3 & 27.3  &  -19.90  & 309  &  3 & 0.058 & 0.083 & 0.69   \\
 349& 138     & 207+026+006 &893   &  5 & 188.0 &    768.8 & 42.6 &   -19.15 & 703  &  5 & 0.056 & 0.119 & 0.47   \\
 351& 138     & 207+028+007 &615   &  4 & 225.4 &    689.1 & 33.0 &   -19.15 & 611  &  4 & 0.058 & 0.088 & 0.65   \\
 362& 158    & 232+029+006  & 306  &  2 & 195.2 &    284.5 & 15.1  &  -18.80  & 305  &  2 &-0.004 &-0.004 &  1.00   \\
 366& 158    & 217+020+010  & 353  &  0 & 300.4 &    763.4 & 31.1  &  -19.90  & 339  &  4 & 0.063 & 0.165 &  0.38   \\
 376& 167    & 255+033+008  & 437  &  2 & 258.7 &    658.0 & 27.9  &  -19.50  & 415  &  3 & 0.058 & 0.029 & 1.03   \\
 474&  76    & 133+029+008  & 389  &  3 & 251.2 &    612.6 & 43.3  &  -19.50  & 377  &  5 & 0.069 & 0.221 & 0.31   \\
 512&  91    & 168+002+007  & 371  &  2 & 227.7 &    410.7 & 26.7  &  -19.40  & 321  &  3 & 0.036 & 0.091 & 0.39   \\
 525& 109    & 177+055+005  & 438  &  4 & 154.2 &    312.6 & 19.3  &  -18.50  & 409  &  3 & 0.017 & 0.020 & 0.84   \\
 530&  -      & 192+062+010  & 333  &  0 & 306.8 &    790.3 & 40.4  &  -20.00   & 316  &  5 & 0.077 & 0.223 & 0.34   \\
 548& 143    & 216+016+005  & 314  &  2 & 158.7 &    227.1 & 14.6  &  -18.50  & 294  &  1 & 0.014 & 0.002 & -      \\
 549&  -      & 214+001+005  & 322  &  1 & 162.5 &    225.8 & 16.8  &  -18.60  & 308  &  1 & 0.015 & 0.027 & -      \\
 550& 154    & 227+007+004  & 459  &  3 & 135.1 &    287.5 & 18.2  &  -18.10  & 447  &  1 & 0.004 & 0.041 & -      \\
 779&  -      & 146+054+004  & 353  &  1 & 139.7 &    209.5 & 20.4  &  -18.00  & 353  &  2 & 0.013 & 0.014 & 0.94   \\
 796&  93    & 167+026+003  & 369  &  1 & 102.3 &    161.7 & 11.9  &  -17.70  & 336  &  1 &-0.003 & 0.001 &  -      \\
 827&  -      & 189+003+008  & 405  &  0 & 254.1 &    572.4 & 30.0  &  -19.50  & 384  &  4 & 0.045 & 0.128 & 0.35   \\
\hline
\end{tabular}\\
\tablefoot{                                                                                 
Columns in the Table are as follows:
(1): $SCl_{ID}$: the supercluster ID in the L10 catalogue;
(2): $ID_{E01}$: the supercluster ID  in  the E01 catalogue; 
(3): the supercluster ID AAA+BBB+ZZZ, where AAA is R.A., +/-BBB is Dec., and CCC is 100$z$;
(4): the number of galaxies in the supercluster, $N_{\mbox{gal}}$;
(5): the number of groups with at least 30 member galaxies in the supercluster, $N_{\mbox{30}}$;
(6): the distance of the density maximum, $d_{\mbox{peak}}$;
(7): the total weighted luminosity of galaxies in the supercluster, $L_{\mbox{tot}}$;
(8): the supercluster diameter (the maximum distance between galaxies in
the supercluster), $\mathrm{Diam}$;
(9): the absolute magnitude limit  of the volume limited sample, $R_{\mbox{magV}}$;
(10): the number of galaxies in the volume limited sample, $N_{\mbox{galV}}$;
(11): the maximum value of the fourth Minkowski functional,
($V_{3,\mathrm{max}}$ (clumpiness), for the supercluster;
(12 -- 14): the shapefinders $K_1$ (planarity) and $K_2$ (filamentarity), and 
the ratio of the shapefinders $K_1/K_2$ 
for the full supercluster (the ratio of the shapefinders is not determined for
superclusters with $V_{3,\mathrm{max}} = 1$ for the full range of threshold densities, as 
explained in text).
}
\end{table*}

We calculated the smoothed luminosity density field of galaxies 
and determined
extended systems of galaxies (superclusters) using this density field. 
To determine superclusters, we created a set of density contours by choosing a 
series of density thresholds. 
We define connected volumes above a certain density threshold 
as superclusters. Different threshold densities correspond to different 
supercluster catalogues. In order to choose proper density levels to determine 
 individual superclusters, we analysed the  density field superclusters at a 
series of density levels. 
The mean luminosity density of our sample is 
$\ell_{\mathrm{mean}}$ = 1.526$\cdot10^{-2}$ $\frac{10^{10} h^{-2} L_\odot}{(\vmh)^3}$.
We chose  the density level $D = 5.0$ 
(in the units of mean density) to 
determine  individual superclusters. At this density level superclusters in 
the richest chains of superclusters in the volume under study  still form 
separate systems; at lower density levels they merge into huge percolating 
systems. At higher threshold density levels superclusters are  smaller and their
number decreases. Details of the calculation of the luminosity density field
and of the supercluster catalogue are
given in Appendix~\ref{sec:DF} and in  \citet{2010arXiv1012.1989J}.

\begin{figure}[ht]
\centering
\resizebox{0.40\textwidth}{!}{\includegraphics*{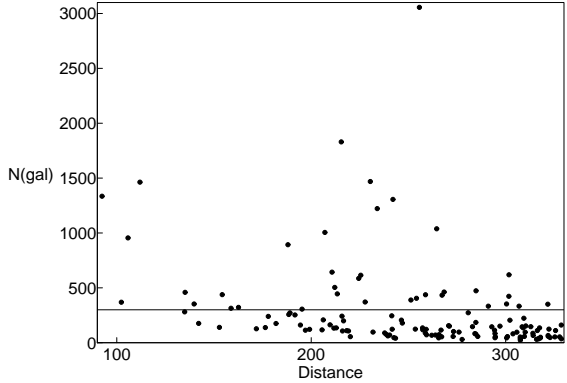}}
\caption{Richness of superclusters vs. their distance (in Mpc/$h$).
The line marks the value $N_{gal} = 300$.
}
\label{fig:scldistngal}
\end{figure}

Figure~\ref{fig:scldistngal} shows the richness of superclusters vs. their distance.  
At distances
less than 200 \Mpc\ there are only a few 
superclusters with less than 300 member galaxies. This selection effect is owed 
to the sample geometry -- it is a pyramid with the top at the location of the observer. At  the distance of 100 \Mpc\ the size of the base 
is 220x140~\Mpc\ only, and it contains only a few superclusters. At 
distances between 100 and 200 \Mpc\ the number of superclusters is small --
there is a void region between the nearby superclusters and those at distances 
larger than 200 \Mpc\ (we will describe the large-scale distribution of superclusters 
in Sect.~4). To avoid these selection effects and to be able to 
resolve the details of supercluster's density distribution (their morphology),
we chose all superclusters in our distance interval with 
at least 300 observed member galaxies; Fig.~\ref{fig:scldistngal} 
shows that they are present at all distances. Data of these superclusters are given in 
Table~\ref{tab:scldata}.
In this table superclusters are ordered as they are presented in 
the text below: first the data of the superclusters with
at least 950 member galaxies, and then of the superclusters with less members.
Throughout the paper we use the 
supercluster ID numbers from the L10 catalogue ($SCl_{ID}$ in 
Table~\ref{tab:scldata}).

To make the calculations of morphology insensitive to selection corrections,
we work with volume-limited samples and the number density of 
galaxies instead of flux-limited samples and luminosity density.
We
recalculated the density field for each individual supercluster 
with a kernel estimator with a $B_3$ box spline as the smoothing kernel, with 
the radius of 8~\Mpc\ \citep[for details we refer to][]{saar06,e07}, and
volume-limited samples of individual superclusters. The absolute 
magnitude limits and  the numbers of galaxies in the volume-limited versions of 
superclusters are given in Table~\ref{tab:scldata}.

\section{Methods}
\label{sect:method}
\subsection{Minkowski functionals and shapefinders}
\label{sect:mink}

The supercluster geometry (morphology) is defined by its outer (limiting) isodensity 
surface and its enclosed volume. The morphology of the isodensity contours is 
(in the sense of global geometry) completely characterised by the four Minkowski 
functionals \mbox{$V_0$ -- $V_3$} (we give the formulas in Appendix 
\ref{sec:MF}).
For a given surface the four Minkowski functionals (from the first to
the fourth) are proportional to the enclosed volume $V$, the area of
the surface $S$, the integrated mean curvature $C$, and the integrated
Gaussian curvature $\chi$.  The last of them, the fourth Minkowski
functional $V_3$,  describes the surface topology; it is a sum of the
number of isolated 
clumps 
and the number of void bubbles minus the
number of tunnels (voids open from both sides) in the region
\citep[see, e.g.][]{saar06}.  High values of the fourth Minkowski
functional $V_3$ suggest a complicated (clumpy) morphology of a
supercluster.

For the argument labelling the isodensity surfaces, we use the (excluded)  mass 
fraction $mf$ -- the ratio of the mass in regions with {\em lower} density than 
at the surface, to the total mass of the supercluster. 
The value $mf=0$ corresponds to 
the whole supercluster, and $mf=1$ to its highest density peak.

With the fourth Minkowski functional $V_3$ we describe the
clumpiness of the galaxy distribution inside superclusters -- the fine
structure of superclusters. 
When the density level is higher than the value used to determine
a supercluster, the
isodensity surfaces move from the outer regions of a supercluster into
its central regions (the value of the mass fraction runs from 0 to 1).
Therefore some galaxies in the outer regions of a supercluster
do not contribute to the
supercluster any more and this changes the inner morphology of the
supercluster, which is reflected in the $V_3 - mf$ relation
(the number of isolated clumps changes, void bubbles, and tunnels may
appear inside superclusters).  We calculate $V_3$ for
superclusters for a range of threshold densities (mass fractions), starting with the
lowest density that determines superclusters ($mf=0$), up to the peak density
in the supercluster core ($mf=1$ß). 
In Section~\ref{sect:morph}
we present figures showing the fourth Minkowski
functional $V_3$ for the whole threshold density interval for each
supercluster  and give the maximum value of $V_3$ for each
supercluster.

\begin{figure*}[ht]
\centering
\resizebox{0.96\textwidth}{!}{\includegraphics*{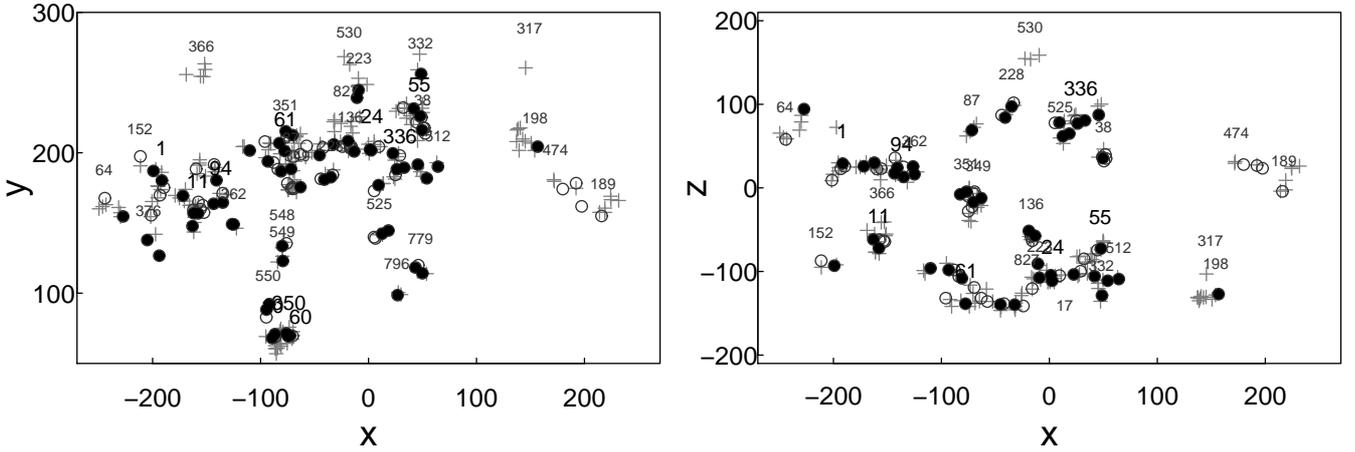}}
\caption{The distribution of groups with at least twelve member galaxies
in our superclusters in cartesian coordinates (see text), in units of \Mpc. 
To make projection effects less significant 
we plot in the right panel a slice with $y > 140$\Mpc. 
The filled circles denote groups with at least 50 member galaxies,
empty circles denote groups with 30--49 member galaxies
and crosses denote groups with 12--29 member galaxies.
The numbers are supercluster ID's (Table~\ref{tab:scldata}, col. 1).
Larger ID numbers show superclusters with at least 950 member galaxies,
smaller  ID numbers -- superclusters with less than 950 member galaxies.
}
\label{fig:sclid}
\end{figure*}

The first three Minkowski functionals have been used 
to calculate the dimensionless 
shapefinders $K_1$ (planarity) and $K_2$ (filamentarity) 
\citep{sah98,sss04,saar09}.  The shapefinders are calculated in two steps. At first, 
specific combinations of Minkowski functionals are used to calculate the 
shapefinders, which describe the thickness, the width, and the length of a supercluster.
With the thickness and the width we calculate the planarity $K_1$
of a supercluster, and with the width and the length we calculate
the filamentarity 
$K_2$ of a supercluster. We give their formulae in Appendix~\ref{sec:MF}.  
\citet{e07}
showed
that in the $(K_1,K_2)$ shapefinder plane the morphology of
superclusters is described by a characteristic curve (morphological
signature).  When  the mass fraction increases, the changes in the
morphological signature accompany the changes of the fourth Minkowski
functional. As the mass fraction increases, 
the planarity
$K_1$ almost does not change for a wide range of mass fractions
(up to $mf \approx 0.35$),  
while the filamentarity $K_2$ increases
-- higher density regions of a supercluster are more filamentary
than the whole supercluster. At higher mass fractions 
the planarity of
superclusters starts to decrease. 
In rich superclusters studied previously
\citep{e07} at a mass fraction of about $mf = 0.7$, the characteristic
morphology of the supercluster changes rapidly. Both the planarity and
filamentarity decrease -- this is a
crossover from the outskirts of
the supercluster to the core of the supercluster. In high-density, clumpy
cores of superclusters, where the isodensity surfaces have a complex shape, 
the planarity and filamentarity may even become
negative \citep{2003MNRAS.343...22S,e08}.
In Fig.~\ref{fig:scl12694} (Sect.~\ref{sect:sclnotes})
we plot the
morphological signature  
with symbols of a size proportional  to the value of the fourth
Minkowski functional $V_3$ at a given mass fraction, to show how the
clumpiness of the supercluster changes together with the change in
the morphological signature.

The morphological signature characterises the 
morphology of superclusters in the  whole threshold density interval
(the fine structure of superclusters).
The values of the shapefinders $K_1$ and $K_2$, 
and  their ratio, $K_1$/$K_2$ (the shape parameter) for the whole supercluster 
quantify the overall shape of superclusters.
The ratio of shapefinders has been used to 
characterise the shape of the whole supercluster, for example,  by 
\citet{kbp02,bas03, 2003MNRAS.343...22S, 2011MNRAS.411.1716C}.

\subsection{Multidimensional normal mixture modelling with {\it Mclust}}
\label{sect:mclust}

We employed multidimensional normal mixture modelling to search for possible subsets 
among superclusters according to their physical and morphological parameters. To 
find an optimal model for the collection of subsets,  
the {\it Mclust} 
package for classification and clustering was applied. 
This package searches for an optimal model for the 
clustering of the data among models with varying shape, orientation and 
volume, finds the optimal number of concentrations,
and the corresponding classification 
(the membership of each concentration).

The {\it Mclust} package gives two statistical measures to estimate  how well  
the superclusters are divided into the subsets.
First, {\it Mclust} calculates the classification uncertainty of
for each object in a dataset.
This parameter is  defined by the probabilities 
for each object to belong to a particular subset and  
is calculated as 1. minus the highest probability of a supercluster
 to belong to a given subset.
The classification uncertainty can be used
as a statistical estimate of  
how well objects are assigned to the subsets.

To measure how well the subsets are determined and to find the best model for a 
given dataset, {\it Mclust} uses the Bayes Information Criterion (BIC), which is based 
on the maximized log-likelihood for the model, the number of variables and the number of 
mixture components. The model with the lowest value of the BIC among all models 
calculated by {\it Mclust} is considered the best. For details we refer to
\citet{fraley2006}. 
Below we calculate both the uncertainities of the classification
of superclusters in the best model determined using {\it Mclust},
and the values of the BIC for different classifications
of superclusters as found by {\it Mclust}.

\section{Results}
\label{sect:results}

\subsection{Large-scale distribution of superclusters}
\label{sect:LSS}

We start with the identification of Abell clusters 
among groups with at least 30 member galaxies
in our superclusters.
With these data we identify our superclusters with those determined
earlier on the basis of Abell clusters (E01). This will help   to analyse 
the large-scale distribution of superclusters and to compare it with 
earlier studies. 

We present a list of the Abell clusters in superclusters in
Table~\ref{tab:sclabell}. 
Here the X-ray clusters are also marked; about 1/3
of the Abell clusters in Table~\ref{tab:sclabell} are X-ray sources.  
These
data were used to compare superclusters with those determined 
in E01. 
We give the E01's ID number if there is at
least one Abell cluster in common between E01 and the present supercluster sample. 
A word of caution is needed -- a common cluster does
not always mean that superclusters can be fully identified with
each other. 
A number of superclusters from E01 are split between several
superclusters in our present catalogue. 
In these cases the identification of  superclusters
 with the superclusters found on the basis of the Abell clusters is
complicated and has to be taken as a suggestion only.

Table~\ref{tab:scldata} and table~\ref{tab:sclabell} show that in our
sample 
there are
three superclusters without any group/cluster with at
least 30 member galaxies, and seven superclusters that do not contain
Abell clusters. Among  the superclusters eight have one group/cluster with
at least 30 member galaxies. All these systems are poor,
comparable with the Local Supercluster, which has only one rich
cluster.  Two of the superclusters contain at
least 10 groups/clusters with at least 30 member galaxies; these are
the richest and most luminous superclusters 
in the sample: SCI 061 and the Corona
Borealis.

\begin{table}[ht]
\caption{Abell clusters in superclusters.}
\label{tab:sclabell}                                                                                    
\begin{tabular}{rrl} 
\hline\hline 
(1)&(2)&(3)\\      
\hline 
 $SCl_{ID}$ &$ID_{E01}$& Abell ID \\ 
    \hline
   1& 162     & 2142x, 2149x   \\ 
  10& 160     &2152    \\                                                 
  11& 154     & 2040, 2028   \\ 
  24& 111     & 1424, 1516  \\
  38&  95     & 1173x, 1187   \\
  55& 111     & 1358  \\
  60& 160     &2197    \\      
  61&126, 136 & 1620, 1650x, 1658x, 1663x, 1692,  \\
   &          & 1750x, 1773x, 1780, 1809x \\
  64& 164     & 2223, 2244  \\
  87&         & 1904   \\         
  94& 158     & 2067, 2065x, 2089  \\
136& 271     & 1569  \\
152& 160     & 2048, 2055x   \\
 198& 82      &  933   \\
 223& 111     &  1541, 1552  \\
 228& 133     & 1767x  \\
 332& 106     &  1346  \\
 336& 109     & 1279, 1436  \\
 349& 138     & 1795x, 1827, 1831x  \\
 350& 160     & 2052x, 2063x  \\
 351& 138     & 1775x, 1800x, 1831x  \\
 362& 158     &  2073, 2079, 2092  \\
 376& 167     &  2249x  \\
 474&  76     &  699  \\
 512&  91     & 1205x, 1238   \\
 525& 109     & 1291x, 1377   \\
 548& 143     & 1913   \\
 550& 154     & 2028, 2055   \\
 796&  93     & 1185x   \\
\hline                                
\end{tabular}\\
\tablefoot{                                                                                 
Columns in the table are as follows:
(1): $SCl_{ID}$: the supercluster ID in the L10 catalogue; 
(2): $ID_{E01}$: the supercluster ID in  the E01 catalogue;  
(3): the Abell ID. x denotes X-ray clusters \citep[E01,][]{boe04}.
}
\end{table}

We show the large-scale distribution of  rich clusters in superclusters in 
Fig.~\ref{fig:sclid}  in cartesian coordinates.
These coordinates are defined as in
\citet{2007ApJ...658..898P,2010arXiv1012.1989J}:
\begin{equation}
\begin{array}{l}
    x = -d \sin\lambda, \nonumber\\[3pt]
    y = d \cos\lambda \cos \eta,\\[3pt]
    z = d \cos\lambda \sin \eta,\nonumber
\end{array}
\label{eq:xyz}
\end{equation}
where $d$ is the comoving distance, and $\lambda$ and $\eta$ are the SDSS survey coordinates.
To complement this figure, we present
in Appendix~\ref{sec:3dfig} a 3D version of Fig.~\ref{fig:sclid},
and  the 3D distributions of
groups in superclusters with the right ascensions, declinations, and
distances of groups.

In Fig.~\ref{fig:sclid} the superclusters SCl~060 and 
SCl~350 (Table~\ref{tab:scldata}) are seen close to us. 
They belong to the Hercules supercluster, which is split
into several superclusters in our sample. These are the nearby rich 
systems seen in Fig.~\ref{fig:scldistngal}.  A chain of 
poor superclusters connects the Hercules supercluster with rich superclusters at 
a distance of about 200~\Mpc\, the superclusters SCl~349 and SCl~351 (the Bootes 
supercluster) among them. 
We  mentioned  in Sect.~2 that   
at small distances the size of the sample cross-section
is only 220x140~\Mpc\, and these superclusters may  be broken up by the sample borders.
Therefore the data on nearby 
superclusters are less reliable 
than the data on the more distant ones.

Rich superclusters at distances of about 210--260~\Mpc\ form three chains, 
separated by voids. A 3D figure on our web pages shows that actually only one 
of these supercluster systems ia a clear chain-like system. This 
is the Sloan Great Wall (SGW), the richest galaxy system in the nearby Universe 
\citep{vogeley04,gott05,nichol06,2010A&A...522A..92E,
2011arXiv1101.1961L, 2011MNRAS.410.1837P, 2011arXiv1105.1632E}. 
The SGW consists of several superclusters of galaxies. 
The richest of them are the superclusters 
SCl~061 and SCl~024. 
The other two supercluster chains are much poorer and cannot 
really be called chains. One of them is separated from the SGW by a 
void; the Bootes supercluster is the richest supercluster in this system. The 
richest supercluster in the third system of superclusters is the Ursa Majoris 
supercluster (SCl~336). 

The very rich Corona Borealis supercluster 
is located at the joint of these systems. This supercluster is a 
member of a huge system of rich superclusters located at the right angle 
with respect to the Local Supercluster, described by \citet{1997A&AS..123..119E}
as the dominant supercluster plane.

 At high positive values of the $x$ coordinate there are no rich
clusters, and the superclusters in this region are also poor. There are some
poor superclusters farther away, perhaps connecting superclusters in
our sample volume with more distant superclusters. 
Thus the large-scale distribution of the superclusters
is very inhomogeneous, as noted also in L10.

To make projection effects less significant, we do not show  the chains of 
nearby superclusters in the right panel of Fig.~\ref{fig:sclid}. The 
superclusters SCl~548, SCl~549, SCl~550, SCl~350, and SCl~060 are superimposed on 
the superclusters SCl~351 and SCl~354, and the superclusters SCl~779 and SCl~796 
on the superclusters SCl~038, SCl~336, and SCl~525.

\subsection{Morphology of superclusters}
\label{sect:morph}

The
results on the morphology of superclusters are summarised
in Table~\ref{tab:scldata} where we list the 
following morphological characteristics for each supercluster: the maximum value of the
fourth Minkowski functional $V_{3,\mathrm{max}}$ (clumpiness), the
values of the shapefinders $K_1$ (planarity) and $K_2$ (filamentarity)
for the whole supercluster, and the ratio of the shapefinders $K_1$/$K_2$
for the whole supercluster. This ratio is not given for the
superclusters  for which $V_3 = 1$ over the whole mass fraction
interval, because the $K_2$ may become very small, making
the ratio $K_1$/$K_2$ noisy.

We began the analysis of the structure of superclusters by searching 
for possible subsets defined 
by their physical and morphological 
characteristics. 
For this purpose  we applied the {\it Mclust} package for classification and 
clustering, 
decribed  shortly in Sect.~\ref{sect:mclust}. 
Initial data for {\it Mclust} was the number of groups with 
at least 30 member galaxies, the total weighted luminosity of galaxies 
in a supercluster,  its diameter, and the morphological 
parameters given in Table~\ref{tab:scldata}. 
The results of this analysis show 
that superclusters can be divided into two main sets.  
The first set consists of superclusters with shape
parameter $K_1/K_2 < 0.6$, and second set of those with shape parameter 
$K_1/K_2 > 0.6$ -- these superclusters are less elongated than the
superclusters in the first set. According to {\it Mclust}, the richest 
supercluster in our sample, the supercluster SCl~061, forms a separate subgroup 
owing to its very high luminosity. For simplicity, we include this supercluster in 
the following analysis in the set of more elongated superclusers.

We estimated how well superclusters are assigned to the two different sets using 
the uncertainity of classification calculated by {\it Mclust} (see 
Sect.~\ref{sect:mclust}). For our superclusters, the mean uncertainty of the 
classification is $1.9\cdot10^{-2}$, showing that superclusters are well 
classified. To measure how well the sets are determined and to find the best 
model of a given dataset we use the BIC values given by {\it Mclust}. For our 
sets of superclusters the lowest value of the BIC corresponds to the division of 
superclusters into two main sets, with SCl~061 forming a separate class. The 
value of the BIC for the one-component model for superclusters is higher, 
showing that this model is less likely for our superclusters.

\begin{table}[ht]
\caption{Parameters of superclusters in sets according to their morphology.}
\label{tab:sclsets}  
\begin{tabular}{lccc} 
\hline\hline 
(1)&(2)&(3)&(4)\\   
        &\multicolumn{2}{c}{$K_1/K_2 < 0.6$} & \multicolumn{1}{c}{$K_1/K_2 \geq 0.6$} \\     
        &\multicolumn{1}{c}{with SCl~061}    & \multicolumn{1}{c}{without SCl~061} & \\     
\hline 
$N_{\mbox{scl}}$& 16  & 15 & 20  \\
$Distance$      & 254 $\pm$ 64  & 254 $\pm$ 66    & 200 $\pm$ 43\\
$L_{\mbox{tot}}$& 820 $\pm$ 316 & 790 $\pm$ 274   & 525 $\pm$ 122  \\
$N_{\mbox{gal}}$& 530 $\pm$ 207 & 442 $\pm$ 176   & 445 $\pm$ 129   \\
$N_{\mbox{30}}$ & 2.5 $\pm$ 0.92 & 2.0 $\pm$ 0.74  & 2.0 $\pm$ 0.63  \\
$Diameter$      & 44  $\pm$ 12  & 43 $\pm$ 11     & 22 $\pm$ 5 \\
$V_{3}$         & 4.5 $\pm$ 1.25 & 4.0 $\pm$ 1.15  & 2.0 $\pm$ 0.48 \\  
$K_{1}$         & 0.08$\pm$ 0.02& 0.08 $\pm$ 0.02 & 0.03 $\pm$ 0.007 \\
$K_{2}$         & 0.20$\pm$ 0.05 & 0.200 $\pm$ 0.05 & 0.031 $\pm$ 0.008 \\

\hline
\end{tabular}\\
\tablefoot{                                                                                 
Columns in the Table are as follows:
1: The parameter: 
 $N_{\mbox{scl}}$, the number of superclusters;
the median distance of superclusters, in Mpc/$h$; 
the median total weighted luminosity of galaxies in the 
superclusters, $L_{\mbox{med}}$, in $10^{10}h^{-2} L_{\sun}$;
the median number of galaxies in the superclusters, $N_{\mbox{gal}}$;
the median number of groups with at least 30 member 
galaxies in the superclusters, $N_{\mbox{30}}$;
the median diameter of the superclusters,  in Mpc/$h$;
the median value of the fourth Minkowski functional, $V_3$;
the median values of the planarity $K_1$ and filamentarity $K_2$
of the superclusters.  
2--3: The median values and $1\sigma$ errors of the corresponding
parameter for  the first set of superclusters with
the ratio of the shapefinders $K_1/K_2 < 0.6$;
4: The median values and $1\sigma$ errors of the corresponding
parameter for  the second set of superclusters with
the ratio of the shapefinders $K_1/K_2 \geq 0.6$.
}
\end{table}

\begin{figure}[ht]
\centering
\resizebox{0.20\textwidth}{!}{\includegraphics*{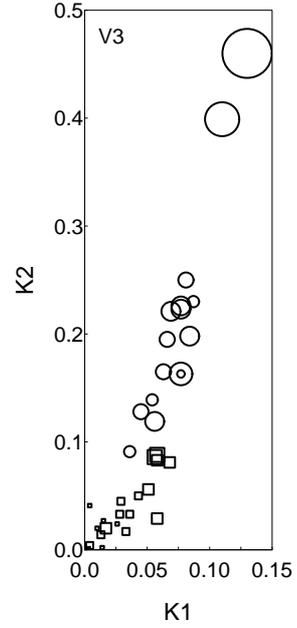}}
\caption{Shapefinders $K_1$ (planarity) and $K_2$ (filamentarity) for
the superclusters with their clumpiness.
The symbol sizes are proportional to the fourth
Minkowski functional $V_{3,\mathrm{max}}$.
Circles denote superclusters with the  shapefinders ratio $K_1/K_2 < 0.6 $
and squares denote the superclusters with the ratio 
$K_1/K_2 \geq 0.6 $.  }
\label{fig:sclk1k2v3dist}
\end{figure}

\begin{figure}[ht]
\centering
\resizebox{0.48\textwidth}{!}{\includegraphics*{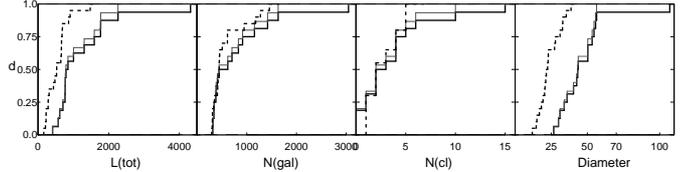}}\\
\caption{Cumulative distributions of the physical characteristics of 
superclusters for the two sets of superclusters
divided by the value of the shape parameter. 
From left to right: the distributions of luminosities, 
numbers of galaxies, numbers of rich groups with at least 30 member
galaxies, and diameters of superclusters.
The black lines correspond to the first set of superclusters
with $K_1/K_2 < 0.6$, including the supercluster SCl~061, the grey lines
to the first set of superclusters without SCl~061, and
the dashed lines  to the second set of superclusters with 
$K_1/K_2 > 0.6$.
}
\label{fig:set12prop}
\end{figure}

\begin{figure}[ht]
\centering
\resizebox{0.45\textwidth}{!}{\includegraphics*{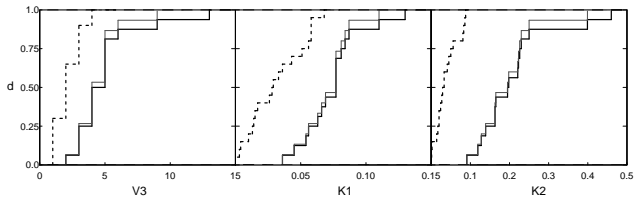}}
\caption{Cumulative distributions of morphological parameters of 
superclusters for two sets of superclusters
divided by the value of the shape parameter. 
From left to right: the distributions of the 
fourth Minkowski functional $V_3$, of planarities $K_1$, and filamentarities
$K_2$ of superclusters.
The black lines correspond to the first set of superclusters
with $K_1/K_2 < 0.6$, including the supercluster SCl~061, the grey lines
to the first set of superclusters without SCl~061, and
the dashed lines to the second set of superclusters with 
$K_1/K_2 > 0.6$.
}
\label{fig:set12morf}
\end{figure}

Fig.~\ref{fig:sclk1k2v3dist} presents 
the shapefinders $K_1$ (planarity) and 
$K_2$ (filamentarity) in the shapefinder's plane for our  supercluster sample. 
In this figure circles correspond to those superclusters for which the ratio of 
the shapefinders $K_1/K_2 < 0.6$, and squares to the superclusters with $K_1/K_2 
\geq 0.6$, i.e. to more elongated and to less elongated superclusters, 
respectively.  The symbol sizes are proportional to the maximum value of the 
fourth Minkowski functional $V_{3,\mathrm{max}}$ (clumpiness). This figure 
reflects both the outer shape and the inner clumpiness (fine structure) of 
superclusters and summarises the morphological information about superclusters. 

Table~\ref{tab:scldata} and  Fig.~\ref{fig:sclk1k2v3dist} 
demonstrate that almost all superclusters in our 
sample are elongated; they have large filamentarities with larger range  
of values than planarities. Two superclusters with the largest 
filamentarities and with the ratio of the shapefinders as small as $K_1/K_2 = 0.28$ 
are the most extreme cases of 
filamentary systems
 in our sample. These two 
superclusters are the richest in our sample: SCI~061 and the Corona Borealis. 
We do not have extremely planar 
superclusters in our sample ($K_1/K_2 \gg 1.0$). Three superclusters 
in our sample have $K_1/K_2 \approx 1.0$ (they are spherical), 
and all these have a small maximum 
clumpiness ($V_{3,\mathrm{max}} \leq 3$).
More elongated superclusters typically have higher values of  clumpiness 
$V_3$ than 
less elongated ones -- they have a more complicated morphology 
(see Table~\ref{tab:sclsets}). 

\begin{figure}[ht]
\centering
\resizebox{0.46\textwidth}{!}{\includegraphics*{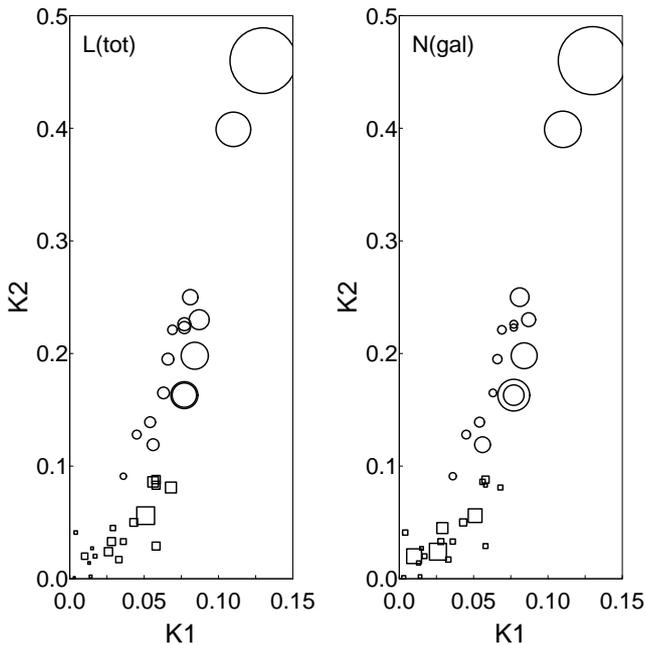}}
\caption{Shapefinders $K_1$ (planarity) and $K_2$ (filamentarity),
 with the luminosities and richnesses of the superclusters.
  In the left panel symbol sizes are proportional to the total
  weighted luminosity of galaxies in the superclusters,  and in the right
  panel to the number of
  galaxies in the superclusters. 
Circles and squares correspond to two sets of superclusters
 as in Fig.~\ref{fig:sclk1k2v3dist}.
 }
\label{fig:sclk1k2lngal}
\end{figure}

\begin{figure}[ht]
\centering
\resizebox{0.40\textwidth}{!}{\includegraphics*{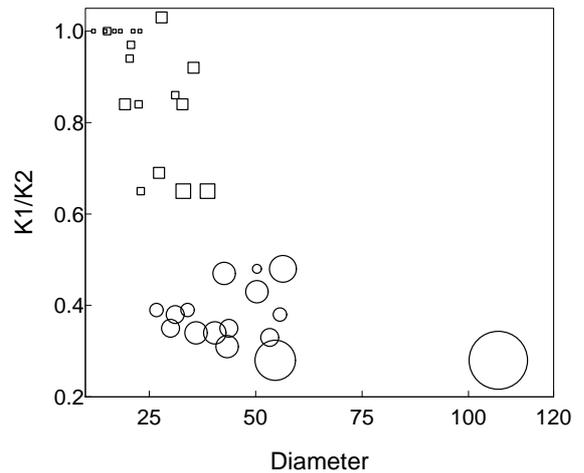}}
\caption{Diameters of superclusters (in Mpc/$h$) vs. their shape parameter. 
Symbol sizes are proportional to the value of the fourth Minkowski
functional $V_3$ of a supercluster. 
Circles and squares correspond to the two sets of superclusters
as in Fig.~\ref{fig:sclk1k2v3dist}.
}
\label{fig:sclk12diam}
\end{figure}

We present the median values of supercluster 
parameters in these sets in Table~\ref{tab:sclsets}. In Fig.~\ref{fig:set12prop} 
we plot the cumulative distributions of the values of the physical characteristics 
of superclusters from the two sets, and in Fig.~\ref{fig:set12morf} the 
cumulative distributions of the
morphological parameters. The scatter of the parameters is large, 
but 
the superclusters in the first set with the shape parameter 
$K_1/K_2 < 0.6$ (with and without the supercluster SCl~061) are richer, more 
luminous, and have larger diameters 
 than  
those 
in the second set with shape 
parameter $K_1/K_2 \geq 0.6$.
Superclusters from the first set also have higher maximum values of 
the fourth Minkowski functional $V_3$, and higher values of planarities
$K_1$ and filamentarities $K_2$.
In Fig.~\ref{fig:sclk1k2lngal} we show another presentation
of these results -- the shapefinder's plane for superclusters 
of different total luminosity and the richness (the number of galaxies in a 
supercluster) as coded in the symbol sizes explained in figure captions. 
The two most 
elongated superclusters are the richest and 
the most luminous.

Figure~\ref{fig:sclk12diam} and Figure~\ref{fig:set12prop}
show  that more elongated superclusters 
also have larger diameters 
than less elongated ones, implying that the systems with
larger diameters are not planar structures. Also, there are no
compact, planar, and very luminous superclusters.

Table~\ref{tab:scldata} shows that all superclusters with the ratio 
of shapefinders with $K_1/K_2 \geq 0.6$ are relatively nearby,
only four of them lie at  distances larger than 250~\Mpc. The closest of them 
is located at about 90~\Mpc, and their median distance is about 
200~\Mpc. As mentioned above, owing to the sample geometry, the nearby superclusters may 
not be fully included in the sample volume and their small 
clumpiness may be due to this selection effect. Some nearby poor superclusters 
are located in low-density filaments between us and more distant superclusters, 
and their shape and small clumpiness may be 
real \citep[as noted also in ][]{1997A&AS..123..119E}. The closest more elongated 
superclusters with the shape parameter $K_1/K_2 < 0.6$ are about 170~\Mpc\ 
away from us, their mean and median distances almost coincide and are about 
254~\Mpc\ (this is approximately the distance to the rich superclusters in the SGW).

\begin{figure*}[ht]
\centering
\resizebox{0.94\textwidth}{!}{\includegraphics*{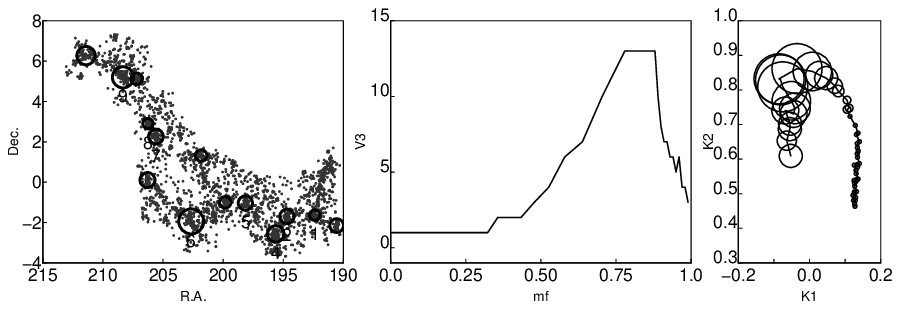}}\\
\resizebox{0.94\textwidth}{!}{\includegraphics*{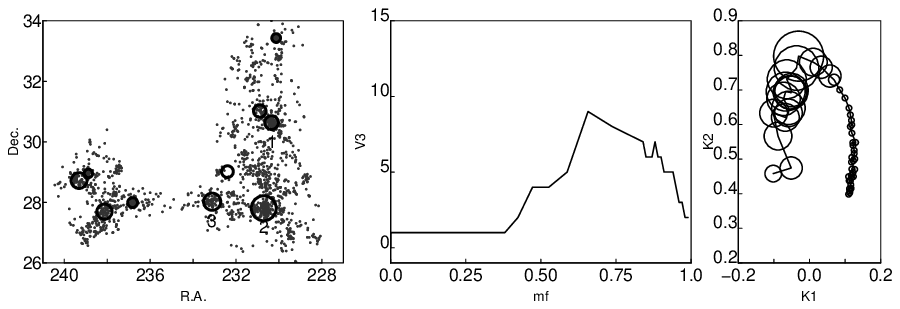}}\\
\caption{Left panels: the distribution of galaxies (grey dots) in the sky, for
the two superclusters, SCl~061 (upper row) and SCl~094 (lower row). Circles mark the location of groups with at least 
30 member galaxies, and the size of a circle is proportional to the size of a group
in the sky. The numbers show Abell clusters.
The middle panels show the fourth Minkowski functional $V_3$, and the right panels 
the shapefinders $K_1$ (planarity) and $K_2$ (filamentarity) 
for a supercluster. The morphological signature in the $K_1-K_2$ plane 
is parametrically defined as $K_1(mf)$ and $K_2(mf)$. 
The sizes of open circles are proportional 
to the value of $V_3$ at a given mass fraction $mf$. They show the change
of the clumpiness with the mass fraction together with the changes in the 
morphological signature. The Abell
clusters in the supercluster SCl~061 (upper row) are 1 -- A1620, 2 -- A1650, 3 -- A1658, 4 -- 1663, 5 --
1682, 6 -- 1750, 7 -- 1773, 8 -- 1780, and 9 -- 1809. The Abell
clusters in the supercluster SCl~094 (lower row) are 1 -- A2067, 2 -- A2065, and
3 -- A2089.  }
\label{fig:scl12694}
\end{figure*}

\subsection{Notes on individual superclusters} 
\label{sect:sclnotes} 

Below, we give a short description of individual superclusters in our sample. 
 In the figures of this section and in Appendix~\ref{sec:MFfig} we show for each 
supercluster (except for those for which $V_3 = 1$ over the whole mass fraction 
interval) the sky distribution of supercluster members, the values of the fourth 
Minkowski functional $V_3$ vs. the mass fraction $mf$ and the morphological 
signature for each supercluster. 
Panels in these figures are as follows. The left panels show  the 
sky distributions of galaxies in superclusters and the location of 
rich groups with at least 30 member galaxies. The middle panels show the 
clumpiness $V_3$ vs. the mass fraction $mf$ for a whole mass fraction interval 
from 0 to 1, and the right panels the shapefinder's $(K_1,K_2)$ curve 
(the morphological signature) for a supercluster. The mass fraction increases 
anti-clockwise along the curves. In these panels  the value of mass 
fraction $mf =0.7$ is marked -- at this value the morphological signature of rich 
superclusters changes.
We will classify our superclusters as spiders, multispiders, filaments, and
multibranching filaments on the basis of their morphological information
and visual appearance. 
Often superclusters are of intermediate type between these main types,
hence for some superclusters our classification is a suggestion only.

We begin with the two richest superclusters from our sample
(Fig.~\ref{fig:scl12694}).  For these two superclusters  the
morphological signature (Fig.~\ref{fig:scl12694}, right panels) is plotted with
symbols of the size proportional to the value of the fourth
Minkowski functional $V_3$ at a given mass fraction, 
to show how the
morphological signature changes together with the
changes in clumpiness, as described in
Sect.~\ref{sect:mink}. 
Here we do not
mark the value of mass fraction $mf=0.7$, for clarity.

{\it The supercluster SCl~061} at a distance of 256~\Mpc\ is the richest member 
of the SGW \citep{2011arXiv1105.1632E, 2011arXiv1101.1961L}. 
This supercluster contains nine Abell clusters,  the 
largest number  in our sample. Five of these are also X-ray clusters 
\citep{boe04}. The richest of them, A1750, is a merging X-ray cluster 
\citep{bel04}. The morphology of  SCl~061 resembles a multibranching filament 
with the maximum value of the fourth Minkowski functional $V_{3,\mathrm{max}} = 
13$, and the ratio of the shapefinders $K_1$/$K_2 = 0.28$, one of the lowest 
in our catalogue (Fig.~\ref{fig:scl12694}, upper row, and 
Table~\ref{tab:scldata}).

{\it The supercluster SCl~094} (the
Corona Borealis supercluster) at a distance of 215~\Mpc\ is the second in
richness among our sample. This system contains three
Abell clusters (A2067, A2065, and A2089), and is a member of the
dominant supercluster plane \citep{1997A&AS..123..119E}. 
The distribution of galaxies in the sky in SCl~094 is plotted in
Fig.~\ref{fig:scl12694} (lower row, left panel). The maximum value of
the fourth Minkowski functional of the Corona Borealis supercluster
$V_{3,\mathrm{max}} = 10$, and the ratio of the shapefinders $K_1$/$K_2 =
0.28$ (Fig.~\ref{fig:scl12694}, the middle and right panels of the lower
row, 
and Table~\ref{tab:scldata}). 
Morphologically SCl~094 is a multispider with a number
of clusters connected by low density filaments, with an overall very
elongated shape that resembles a horse-shoe, with the merging X-ray
cluster A2065 at the top \citep{2006ApJ...643..751C}.  SCl~094
has been studied by \citet{1998ApJ...492...45S}, who found
that the core of this system probably has started to
collapse. Numerical simulations show that such collapsing cores in
superclusters are rare \citep{2002MNRAS.337.1417G}.
\citet{2011arXiv1101.1961L} proposed that SCl~094 may merge 
with several surrounding superclusters in the future.
In last years interest in the Corona Borealis supercluster 
region has grown because of the discovery of the CMB cold spot in it's
direction. This may be partly caused by the warm-hot diffuse gas 
in the supercluster filaments between the clusters, or by some undiscovered distant 
cluster \citep[][and references 
therein]{2008MNRAS.391.1127G,2009MNRAS.396...53P, 
2010MNRAS.403.1531G,2010hsa5.conf..329P}.
{\it The poor superclusters SCl~362 and SCl~366} are also members of the Corona 
Borealis supercluster. The morphology of   SCl~362 resembles a 
simple spider with the ratio of the shapefinders $K_1$/$K_2 = 1.0$, while that 
of  SCl~366 resembles a multibranching filament with the maximum 
value of the fourth Minkowski functional $V_{3,\mathrm{max}} = 4$, and the ratio 
of the shapefinders $K_1$/$K_2 = 0.38$ (Fig.~\ref{fig:sclapp2}).

\begin{figure*}[ht]
\centering
\resizebox{0.94\textwidth}{!}{\includegraphics*{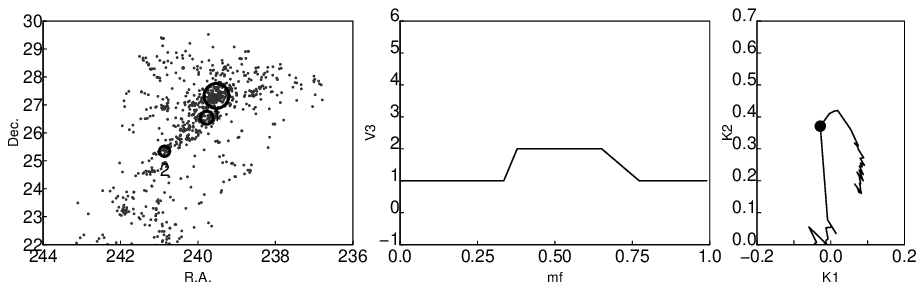}}\\
\resizebox{0.94\textwidth}{!}{\includegraphics*{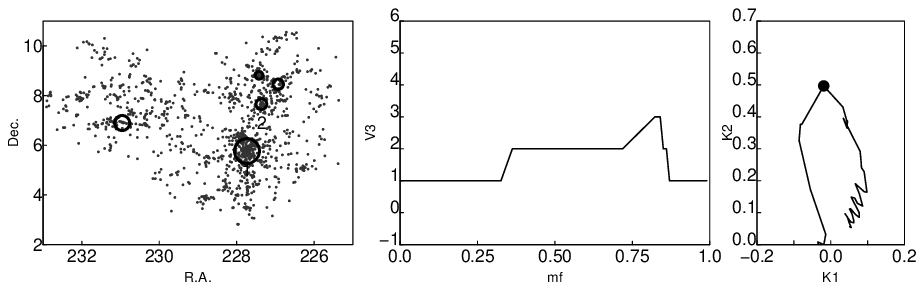}}\\
\resizebox{0.94\textwidth}{!}{\includegraphics*{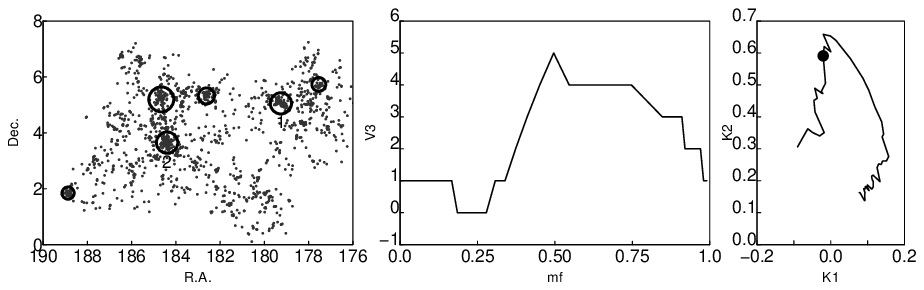}}\\
\caption{Panels as in Fig.~\ref{fig:scl12694}.
Filled circles in the right panel mark the value of the mass fraction 
$mf= 0.7$. 
Upper row: the supercluster SCl~001. The Abell clusters are 1 -- A2142 and 2 -- A2149.
Middle row: the supercluster SCl~011. The Abell clusters are 1 -- A2040 and 2 -- A2028.
Lower row: the supercluster SCl~024. The Abell clusters are 1 -- A1424 and 2 -- A1516.
}
\label{fig:scl001}
\end{figure*}

{\it The supercluster SCl~001}  contains 
two rich Abell clusters, A2142 and A2149, both of them 
are X-ray sources (Table~\ref{tab:sclabell}). Chandra 
observations have revealed that A2142 is  probably still merging 
\citep{2000ApJ...541..542M}. SCl~001 is located in a region with a dense 
concentration of superclusters, close to 
SCl~011 and SCl~094. All these systems form a part of the 
dominant supercluster plane. At the location of
SCl~001  the luminosity density is the highest in the whole SDSS 
survey; this is probably at least partly because of the rich X-ray cluster A2142. 
The morphology of SCl~001 resembles a filament where clusters are 
located almost along a straight line (the 3D model on our web pages shows this best). 
The maximum value of the fourth Minkowski functional for SCl~001 
$V_{3,\mathrm{max}} = 2$ and the ratio of the shapefinders $K_1$/$K_2 = 0.48$ 
(Fig.~\ref{fig:scl001} and Table~\ref{tab:scldata}).

{\it The supercluster SCl~011} at a distance of 234~\Mpc\ contains two Abell 
clusters, A2040  and 
A2028, which are members of different superclusters in E01.  
The maximum value of the fourth Minkowski 
functional $V_{3,\mathrm{max}} = 3$ (Fig.~\ref{fig:scl001}, middle row), 
the ratio of the
shapefinders $K_1$/$K_2 = 0.92$. The morphology of SCl~011 resembles 
a sparse multispider or multibranching filament with a quite uniform density (as 
suggested by the low values of the fourth Minkowski functional $V_3$ for a wide mass 
fraction interval). SCl~011 belongs to the same supercluster complex 
as SCl~001 (Fig.~\ref{fig:sclid}).

{\it The supercluster SCl~024}  at a distance of
230~\Mpc\ is the second richest member of the SGW. 
SCl~024 contains two Abell clusters, A1424 
and A1516. Its morphology resembles a multispider with 
the maximum value of the fourth Minkowski functional $V_{3,\mathrm{max}} = 5$ and
the ratio of the shapefinders $K_1$/$K_2 = 0.48$
(Fig.~\ref{fig:scl001}, lower row). The values of the fourth Minkowski functional 
of SCl~024 at small mass fractions ($mf \approx 0.25$) $V_3 = 0$ 
suggest that SCl~024 has low-density tunnels inside
\citep{2011arXiv1105.1632E}.
The  member of the supercluster SCl~111 in E01 to which SCl~024 also belong, 
the poor supercluster SCl~223, can be described as a multispider with two
concentrations --the fourth Minkowski functional has a
value of 2 for a wide mass fraction interval (Fig.~\ref{fig:sclapp1}).
SCl~223 is separated from  SCl~024 and
SCl~055 by a small void, showing that clusters gathered together into
one supercluster in the E01 catalogue sometimes do not belong to the same
supercluster, when systems are determined using data on galaxies.

\begin{figure*}[ht]
\centering
\resizebox{0.94\textwidth}{!}{\includegraphics*{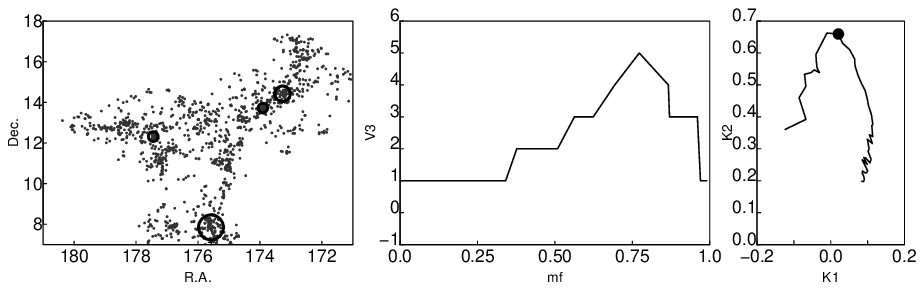}}\\
\resizebox{0.94\textwidth}{!}{\includegraphics*{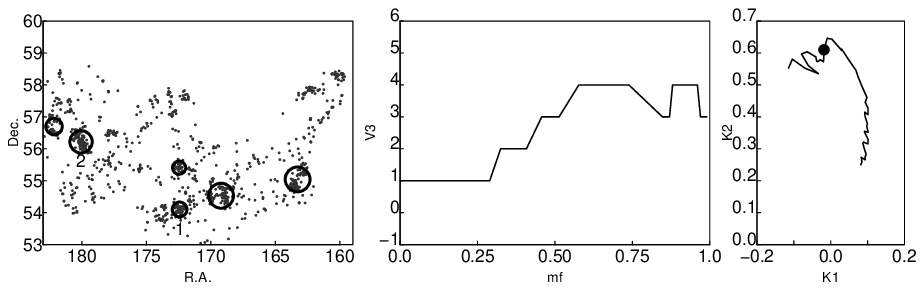}}\\
\resizebox{0.94\textwidth}{!}{\includegraphics*{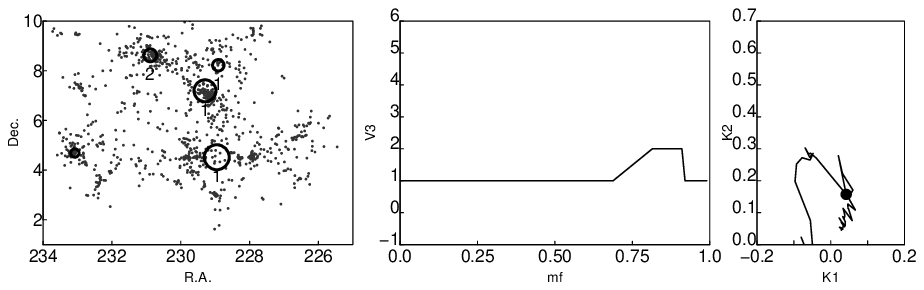}}\\
\caption{Panels are the same as in Fig.~\ref{fig:scl12694}.
Upper row: the supercluster SCl~055. The Abell cluster is A1358.
Middle row: the supercluster SCl~336. The Abell clusters are 1 -- A1279, 2 -- A1436.
Lower row: the supercluster SCl~350. The Abell clusters are 1 -- A2052, 2 -- A2063.
}
\label{fig:scl055}
\end{figure*}

{\it The supercluster SCl~055}  at a
distance of 242~\Mpc\  is
separated from the SGW by a void and is connected 
to it
by a filament of galaxies (this is the reason why 
SCl~024 and SCl~055 both belong to the supercluster SCl~111 in the E01
catalogue). SCl~055 contains one Abell cluster, A1358. The
morphology of SCl~055 resembles a multibranching filament or
an elongated multispider with the maximum value of the fourth
Minkowski functional $V_{3,\mathrm{max}} = 5$ and the ratio of the
shapefinders $K_1$/$K_2 = 0.43$ (Fig.~\ref{fig:scl055}).

{\it The supercluster SCl~336}  contains two Abell clusters, A1279 and A1436, 
which are members of the Ursa Majoris 
supercluster (Fig.~\ref{fig:scl055}, middle row). SCl~336 belongs to a 
chain of superclusters that is separated by a void from  SCl~349 
and SCl~351 (the Bootes supercluster) \citep[see also][and 
references therein]{2009AstBu..64....1K}. \citet{2011arXiv1101.1961L} found that 
several filamentary systems may be associated with this system. According 
to our calculations, the morphology of the Ursa Majoris supercluster resembles a 
sparse multibranching filament with the maximum value of the fourth Minkowski 
functional $V_{3,\mathrm{max}} = 4$,  and the ratio of the shapefinders $K_1$/$K_2 = 
0.33$ (Fig.~\ref{fig:scl055}). 
{\it The poor supercluster SCl~525} is also a member of the Ursa Majoris supercluster,
morphologically SCl~525 can be described as a spider.

{\it The superclusters SCl~010, SCl~060, and SCl~350} belong to the rich Hercules 
supercluster  that is split between several systems in our 
present catalogue. They are located at a distance of about 
100~\Mpc\ and contain clusters that are exceptionally rich in the T10
catalogue, owing to their small distance from us (A2152 in SCl~010 and 
A2197 in SCl~060). This is the reason why the value of the fourth Minkowski 
functional for SCl~010 and SCl~060 $V_3 = 1$ over the whole 
mass fraction interval, showing that they contain only one 
high density clump.  SCl~350 contains five rich clusters, 
three of them correspond to the Abell cluster A2052. This suggests that 
A2052 has a substructure, with different components corresponding to 
different groups and clusters in the T10 catalogue. The maximum value of 
the fourth Minkowski functional for SCl~350, $V_{3,\mathrm{max}} = 2$ 
(Fig.~\ref{fig:scl055}, lower row), and the ratio of the shapefinders $K_1$/$K_2 
= 0.65$. At the farther end SCl~350 
chain joins the dominant supercluster plane (Fig.~\ref{fig:sclid}).

We summarise the results of the morphological analysis of the
superclusters 
in Table~\ref{tab:sclothermorf}. 
For clarity we present the figures of the distribution of galaxies and 
rich groups/clusters  in the sky for  superclusters
with less than 950 members galaxies, as well as their 
fourth Minkowski functionals and the morphological signatures in  
Appendix~\ref{sec:MFfig}.

\begin{table*}[ht]
\caption{Data of the morphology of superclusters.}
\label{tab:sclothermorf}  
\begin{tabular}{rrrrrrrrr} 
\hline\hline 
(1)&(2)&(3)&(4)&(5)& (6)&(7)&(8)& (9)\\      
\hline 
 $SCl_{ID}$ & $d_{\mbox{peak}}$ & $N_{\mbox{gal}}$ &$N_{\mbox{30}}$ 
 & $N_{\mbox{Abell}}$ & $V_{3,\mathrm{max}}$ & Set &$Morph.$& Notes \\
   &  Mpc/$h$&   &   & & & && \\
\hline
   1&   264.5 &1038  &  4 & 2 &    2 & 1 & F &     \\ 
  10&   111.9 &1463  &  2 & 1 &    1 & 2 & S &        \\    
  11&   235.0 &1222  &  5 & 2 &    3 & 2 & mS(mF) &     \\ 
  24&   230.4 &1469  &  4 & 2 &    6 & 1 & mS & SGW     \\
  55&   242.0 &1306  &  4 & 1 &    5 & 1 & mS(mF) &    \\
  60&    92.0 &1335  &  3 & 1 &    1 & 2 & S  &        \\    
  61&   255.6 &3056  & 15 & 9 &   13 & 1 & mF & SGW       \\
  94&   215.4 &1830  & 10 & 3 &    9 & 1 & mS &       \\
 336&   170.0 &1005  &  6 & 2 &    4 & 1 & mF &        \\
 350&   105.8 & 955  &  5 & 2 &    2 & 2 & mS &        \\
\hline
  38&   224.4 &586   &  5 & 2 &    2 & 2 & S  &  \\
  64&   301.7 &619   &  2 & 2 &    3 & 1 & F  &  \\
  87&   213.4 & 445  &  1 & 1 &    2 & 2 & S  &  \\ 
 136&   212.1 &504   &  5 & 1 &    2 & 2 & S  &   \\
 152&   301.6 & 423  &  2 & 2 &    3 & 2 & F  &  \\
 189&   267.2 & 433  &  1 & 0 &    4 & 1 & F  &  \\
 198&   284.7 & 473  &  1 & 1 &    4 & 2 & mF  &  \\
 223&   268.3 & 462  &  2 & 2 &    3 & 1 & mS  &  \\
 228&   210.6 &643   &  4 & 1 &    2 & 2 & F  &  \\
 317&   321.6 & 351  &  1 & 0 &    5 & 1 & mS  & T \\
 332&   291.0 & 333  &  1 & 1 &    3 & 2 & mS  &  \\
 349&   188.0 &893   &  5 & 3 &    5 & 1 & mS  & F(L11)  \\
 351&   225.4 &615   &  4 & 3 &    4 & 2 & mF  &  \\
 362&   195.2 & 306  &  2 & 3 &    2 & 2 & mS  &   \\
 366&   300.4 & 353  &  0 & 0 &    4 & 1 & mF  &   \\
 376&   258.7 & 437  &  2 & 1 &    3 & 2 & F   & T \\
 474&   251.2 & 389  &  3 & 1 &    5 & 1 &  F  &  \\
 512&   227.7 & 371  &  2 & 2 &    3 & 1 & mF  &  \\
 525&   154.2 & 438  &  4 & 2 &    3 & 2 & mS  &  \\
 530&   306.8 & 333  &  0 & 0 &    5 & 1 & mS  & T \\
 548&   158.7 & 314  &  2 & 1 &    1 & 2 & S  & Her-DSP \\
 549&   162.5 & 322  &  1 & 0 &    1 & 2 & S  & Her-DSP \\
 550&   135.1 & 459  &  3 & 2 &    1 & 2 & S  & Her-DSP \\
 779&   139.7 & 353  &  1 & 0 &    2 & 2 & mS  & SGW chain \\
 796&   102.3 & 369  &  1 & 1 &    1 & 2 & S  &  SGW chain \\
 827&   254.1 & 405  &  0 & 0 &    4 & 1 & F  &  \\
\hline
\end{tabular}\\
\tablefoot{                                                                                 
Columns in the Table are as follows:
(1): $SCl_{ID}$: the supercluster ID in the L10 catalogue;
(2): the distance of the density maximum, $d_{\mbox{peak}}$; 
(3): the number of galaxies in the supercluster, $N_{\mbox{gal}}$;
(4): the number of groups with at least 30 member galaxies in the supercluster, $N_{\mbox{30}}$;
(5): the number of Abell clusters 
among groups with at least 30 member galaxies 
in the supercluster, $N_{\mbox{Abell}}$;
(6): the maximum value of the fourth Minkowski functional,
($V_{3,\mathrm{max}}$ (clumpiness), for the supercluster;
(7): classification: set -- 1 for superclusters 
with the ratio of the shapefinders $K_1/K_2 < 0.6$
and 2 for superclusters with the ratio of the shapefinders $K_1/K_2  \geq 0.6$);
(8): morphology: S denotes spiders, mS -- multispiders, F -- filaments and
mF -- multibranching filaments;
(9): notes. SGW denotes members of the Sloan Great Wall;
F(L11) denotes a filamentary system according to
\citep{2011arXiv1101.1961L};
T denotes superclusters for which at a certain mass fraction
 $V_3 = 0$, suggesting that these superclusters have tunnels
through them  
\citep[App.~\ref{sec:MF},][]{2011arXiv1105.1632E}.
 Her-DSP denotes the superclusters located in a chain of 
superclusters that connects the Hercules supercluster
with the dominant supercluster plane, and
SGW chain denotes the superclusters in a chain in the foreground of the
Sloan Great Wall. 
}
\end{table*}

In Table~\ref{tab:sclothermorf} we list for every supercluster
their distance, 
the numbers of galaxies, the number of 
rich groups with at least 30 member galaxies,
and the number of Abell clusters among them. 
From morphological parameters we list the maximum value of the 
fourth Minkowski functional and the number that
shows whether the supercluster belongs to the set 1 (more
elongated superclusters) or to the set 2 (less elongated superclusters).
We also give morphological descriptions of superclusters and notes.

Table~\ref{tab:sclothermorf} shows that superclusters with a smaller number
of galaxies also contain, as expected,  a smaller number of rich groups and Abell
clusters. 
Among them less elongated superclusters 
dominate over more elongated superclusters -- there are 16 systems
from  set 2 and 10 from  set 1 among them. The morphology
of 14 of them can be described as simple spider or simple filament, and
12 of them are either multispiders or multibranching filaments.
In contrast, of 10 superclusters with at least 950 member
galaxies 7 can be described as multispiders of multibranching 
filaments, two of them are simple spiders and one is a
simple filament.

\section{Discussion}
\label{sect:discussion}    
                                                      
\subsection{Selection effects  }
\label{sect:sel}    

The main selection effect in our study comes from the use of the flux-limited 
sample of galaxies to determine the luminosity density field and superclusters.
To keep the luminosity-dependent selection effects as small as possible, we used
  data 
on galaxies and galaxy systems for a distance interval
90--320~\Mpc. In this interval 
these effects are the smallest (we refer to T10 for details). 
We calculate Minkowski functionals  of individual superclusters
from volume limited samples.                           
This approach makes the
calculations of morphology insensitive to luminosity dependent selection 
effects. 

Another selection effect comes from the choice of the density level used to 
determine superclusters. At the density level  applied in the present paper ($D 
= 5.0$), rich superclusters do not percolate yet. For example, in the SGW 
we see several individual rich superclusters. At a lower threshold 
density these superclusters join into a huge system. In the same time the 
Corona Borealis supercluster is split into several 
superclusters in our catalogue at the 
density level  $D = 5.0$ . Thus there is no unique density level, which 
would be the best choice for all superclusters. However, L10 show that 
superclusters are fairly well-defined systems and do not change much when 
changing the density level, if only this change is small and does not break them up
into small systems or does not join them into huge percolating systems. 
When we move towards higher density levels, some galaxies and galaxy systems
in the lower density outskirts of a supercluster do not belong to the supercluster
any more; in that case the clumpiness of the supercluster at low mass fractions
may decrease. Moving towards lower density levels, new galaxies and galaxy
systems in the outskirts of a supercluster 
become supercluster members. This may increase the value of the clumpiness
and change the morphological signature of a supercluster
at low mass fractions. The possible change in morphology depends
on the number of galaxies and on the richness of groups removed
from or added to  the supercluster and is individual for each
supercluster.

At small distances our sample volume is small, which introduces another
selection effect -- nearby rich superclusters are split between
several superclusters, some parts of these superclusters may be located
outside of the sample volume. This makes the data about nearby superclusters
less reliable.

The identification of Abell clusters may also be affected by selection
effects. It is well known that some Abell clusters consist of
several line-of-sight components. A good example of such a cluster
is the cluster Abell 1386, recently studied by \citet{2011MNRAS.410.1837P}.
Our search discarded Abell clusters affected by this projection effect.

\subsection{Comparison with other studies}
\label{sect:others}

Recently
\citet{2011MNRAS.411.1716C} used volume-limited samples of galaxies from the 
SDSS DR7 to extract superclusters of galaxies and to study the morphology of  
whole superclusters with the shape parameter (the ratio of the shapefinders 
$K_1/K_2$).  To determine superclusters the authors calculated the density field 
with an Epanechnikov kernel and found systems of galaxies with at least 10 member 
galaxies. In \citet{e07}, we compared the Epanechnikov and  $B_3$ box spline kernels
and found that both kernels are good to describe the overall shape of 
superclusters, while the $B_3$ box spline kernel better resolves the inner 
structure of superclusters. This is the reason why we used this kernel in 
the present study.  The most important difference between our studies is that we used 
the data about the richer superclusters with at least 300 member galaxies. 
\citet{2011MNRAS.411.1716C} showed that there are both planar and filament-like 
superclusters (pancakes and filaments) among the superclusters of their sample, 
while in our sample there are almost no superclusters for which planarity is 
larger than filamentarity. However, \citet{2011MNRAS.411.1716C} 
found that very rich and luminous 
superclusters tend to be filaments, as we also found in our study.

The overall shapes of superclusters, described by 
the shape parameters or approximated by triaxial ellipses, have been analysed in 
\citet{1998A&A...336...35J, kbp02, bas03, 2007A&A...462..397E, 
2011MNRAS.411.1716C, 2011arXiv1101.1961L}. These studies 
showed that elongated, prolate structures dominate among superclusters, as we 
also found in our study. 
In the analysis of the geometry of the structures around the 
galaxy clusters from simulations
\citet{2011MNRAS.tmp..158N}  showed that these structures tend to lie on 
planes. Similarly, in our study and in the study by \citet{2011MNRAS.411.1716C} 
poorer systems are more planar.
\citet{2003MNRAS.343...22S} used the shapefinders plane 
$K_1$-$K_2$ (the morphological signature in our study)  to study the morphology 
of superclusters in simulations. They found that  richer and more massive 
superclusters tend to be more filamentary and have a more complicated inner 
structure (high values of the genus in their study).

Future evolution of the structure in the Universe has been addressed in 
simulations by several authors, we refer to \citet{2009MNRAS.399...97A} for a 
review and references.  \citet{2009MNRAS.399...97A} studied the evolution of the 
shape and inner structure of superclusters in simulations from the present time 
to a distant future (from $a = 1$ to $a = 100$, 
$a$ is the expansion factor). 
In their study superclusters were defined as high-mass bound objects, and the 
 supercluster shape  was approximated by triaxial ellipses. To analyse the 
substructure of superclusters they used the multiplicity function of clusters in 
superclusters. \citet{2009MNRAS.399...97A} showed that superclusters are 
elongated, prolate structures, there are no thin pancakes among them. Future 
superclusters are typically much more spherical than present-day superclusters. 
Presently, the superclusters contain a large number of clusters, which may merge 
into a single cluster in the far future, i.e., multispiders and multibranching 
filaments may evolve into simple spiders and filaments.

Comparisons of the properties of rich and poor superclusters have 
revealed several  
differences between them. The mean and maximum number densities of 
galaxies in rich superclusters are higher than in poor superclusters. Rich 
superclusters are more asymmetrical than poor superclusters 
\citep{2007A&A...462..397E}. Rich superclusters contain  high-density cores 
\citep{e07b}. The fraction of rich clusters and X-ray clusters in rich 
superclusters is larger than in poor superclusters 
\citep{e2001,2011arXiv1101.1961L}, and the core regions of the richest 
superclusters may contain merging X-ray clusters 
\citep{2000MNRAS.312..540B,rose02}.  However, we still lack
a detailed analysis of whether 
the differences between the properties of the galaxy and group content of rich and poor 
superclusters are related also to the differences in morphology of superclusters.

In \citet{e08} we compared the properties of the two richest superclusters from 
the 2dF Galaxy Redshift Survey, the superclusters SCl~126 (SCl~061 in the 
present study), and the Sculptor supercluster (SCl~9 in E01).  We used Minkowski 
functionals to quantify the fine structure of these superclusters as traced by 
different galaxy populations. Our calculations showed that in the supercluster 
SCl~126 the population of red, early type galaxies is more clumpy than the 
population of blue, late type galaxies, especially in the outskirts of the 
supercluster. In contrast, in the supercluster SCl~9 the clumpiness  of galaxies 
of different type is quite similar in its outskirts. In the core of the 
supercluster SCl~9 the clumpiness of blue, late type galaxies is larger than the 
clumpiness of red, early type galaxies. In the supercluster SCl~111 in the SGW 
the clumpiness of red galaxies is larger than that of blue galaxies 
\citep{2011arXiv1105.1632E}. Morphologically the  supercluster SCl~126 resembles 
a multibranching filament, while the Sculptor supercluster and SCl~111 resemble 
a  multispider. We need to study the morphology and galaxy populations of a 
larger sample of superclusters to find out whether the differences between 
galaxy populations in superclusters are also related to their different 
morphology.

\citet{2010MNRAS.tmp.1270A} recently studied the structure and morphologies of 
elements that define the cosmic web with the Multiscale Morphology Filter and 
data from $\Lambda$CDM simulations. The authors found several typical morphologies for 
filaments, which they describe as line, grid, star, and complex filaments. We can 
compare that with our classification of supercluster morphologies. 
Approximately, line filaments are comparable with our simple filaments. Star 
filaments can be compared with simple spiders. Grid and complex filaments  
may correspond either to multibranching filaments or to multispiders, although 
complex filaments are more similar to multispiders and grid filaments to 
multibranching filaments. This shows that the morphology of observed and 
simulated superclusters is, in general, similar, as we  found earlier from much 
smaller samples of observed and simulated superclusters. The biggest exception 
is the supercluster 
SCl~061, 
a very rich and high-density multibranching 
filament. No other supercluster with such a morphology has been found yet 
either in simulations  in observations \citep[see also][]{gott08}. 
Another difference between the observed and simulated superclusters, as 
quantified by Minkowski functionals and shapefinders, is  fine 
structure of superclusters, delineated by galaxies of different luminosity
 \citep{e07}. The clumpiness of observed superclusters for galaxies of 
different luminosity has a much larger scatter than that of simulated 
superclusters. Simulations do not yet explain all the features 
of observed superclusters.

\section{Summary and conclusions} 
\label{sect:conc}

We have presented an analysis of the large-scale distribution and morphology of 
superclusters from the SDSS DR7. While the overall shape of  superclusters has 
been analysed earlier in several studies, our paper is the first in which the 
inner morphology of a large sample of observed superclusters is studied in 
detail.  We used multidimensional normal mixture modelling  to divide 
superclusters into two sets according to their physical and morphological 
properties. We present 2D and 3D distributions of galaxies and rich groups in 
superclusters, the clumpiness curve (the fourth Minkowski functional $V_3 - mf$ 
relation), as well as the morphological signature $K_1$-$K_2$ for each supercluster in 
our sample . 

The superclusters were selected to 
contain at least 300 observed galaxies. As we showed, this does not mean that 
all superclusters in our sample are rich -- about 1/4 of our superclusters 
contain one rich cluster or group only (three superclusters do not contain any 
rich group or cluster).

Summarising, our study showed that:

\begin{itemize}

\item[1)]
Most superclusters contain Abell clusters, about 1/3 of them 
are X-ray clusters. 

\item[2)] The large-scale distribution of the superclusters is very inhomogeneous. 
Rich superclusters in the sample form three chains, the Sloan Great Wall
is the richest of them.

\item[3)] Almost all  superclusters under study are elongated and
have filamentarities that are 
larger than than their planarities. More elongated superclusters are also more 
luminous, have larger diameters and contain a larger number of rich clusters. 
The values of the fourth Minkowski functional $V_3$ show that they also have a
more complicated inner morphology than less elongated superclusters.

\item[4)] The morphological analysis shows a large variety of morphologies 
among superclusters. The fine structure of superclusters can be described with four main 
types of morphology: spiders, multispiders, filaments, and multibranching 
filaments.  Often a supercluster has an intermediate morphology between these 
main types. 
Superclusters with a similar shape parameter may have a different fine
structure. Consequently neither the 
shape parameter or the number of rich clusters in a supercluster alone 
are sufficient to 
describe the morphology of superclusters.

\end{itemize}

Our results on the 
morphology of superclusters can be used to compare the properties of local and 
high-redshift superclusters. 
Few superclusters at very high redshifts have 
already been discovered in deep, wide-field imaging surveys 
\citep{2005MNRAS.357.1357N, 2007MNRAS.379.1343S, 2008ApJ...684..933G, tanaka09,
2011arXiv1102.4617S}. 
Deep surveys like the ALHAMBRA project \citep{moles08} will provide us with data 
about (possible) very distant superclusters; we can analyse their structure and compare that with the local superclusters, using morphological methods.

Our study does not give a definite answer to the question about the possible 
connection between the morphology of superclusters and their large-scale 
distribution. Also, there are superclusters in our sample that can be 
described as filaments or multibranching filaments, but none of them is as 
rich and has such an overall high density as the supercluster SCl~061. Even 
the richest supercluster with a multispider morphology in our sample, the 
supercluster SCl~094, is not as rich as the very rich Sculptor supercluster in 
\citet{e07}. This shows the need for a larger sample of superclusters 
to understand the morphological variety of 
superclusters, and to study the possible connection between the large-scale 
distribution of superclusters and their morphology.

Different morphologies of superclusters suggest that their evolution has been 
different. To understand the formation and  evolution of superclusters 
of different morphology better, we need to study the properties and evolution of 
superclusters in simulations. 
The morphology of superclusters and its evolution 
may be one of the factors to distinguish between different cosmological models 
\citep{kbp02,2007JCAP...10..016H}.  Especially interesting is the supercluster 
SCl~061 in the Sloan Great Wall. Up to now simulations have not 
been able to model its morphology \citep[][and references therein]{e07}. New 
simulations with larger volumes are needed to study the morphology of 
superclusters and the evolution of the morphology of simulated superclusters, to 
understand the reasons for the exceptional morphology of the supercluster 
SCl~061. In addition, very rich superclusters are rare; 
another reason to use simulations for large volumes comes 
from the demand to include as large a variety of superclusters in the simulation 
volume as possible.

\begin{acknowledgements}
  
We thank our referee for a very detailed review with many useful 
  suggestions that helped to improve the paper. 
  
  Funding for the Sloan Digital Sky Survey (SDSS) and SDSS-II has been
  provided by
  the National Science Foundation, the U.S.  Department of Energy, the
  National Aeronautics and Space Administration, the Japanese Monbukagakusho,
  and the Max Planck Society, and the Higher Education Funding Council for
  England.  The SDSS Web site is http://www.sdss.org/.
  
  The SDSS is managed by the Astrophysical Research Consortium (ARC) for the
  Participating Institutions.  The Participating Institutions are the American
  Museum of Natural History, Astrophysical Institute Potsdam, University of
  Basel, University of Cambridge, Case Western Reserve University, The
  University of Chicago, Drexel University, Fermilab, the Institute for
  Advanced Study, the Japan Participation Group, The Johns Hopkins University,
  the Joint Institute for Nuclear Astrophysics, the Kavli Institute for
  Particle Astrophysics and Cosmology, the Korean Scientist Group, the Chinese
  Academy of Sciences (LAMOST), Los Alamos National Laboratory, the
  Max-Planck-Institute for Astronomy (MPIA), the Max-Planck-Institute for
  Astrophysics (MPA), New Mexico State University, Ohio State University,
  University of Pittsburgh, University of Portsmouth, Princeton University,
  the United States Naval Observatory, and the University of Washington.
  
  We acknowledge the
  Estonian Science Foundation for support under grants No.  8005 and
  7146, and the Estonian Ministry for Education and Science support by grant
  SF0060067s08.  
This work has also been supported by
the University of Valencia through a visiting professorship for Enn Saar and
by the Spanish MEC project AYA2006-14056, ``PAU'' (CSD2007-00060), including
FEDER contributions,  the Generalitat Valenciana project of excellence 
PROMETEO/2009/064, and by Finnish Academy funding.
J.E.  thanks
Astrophysikalisches Institut Potsdam (using DFG-grant Mu 1020/15-1), 
where part of this study was performed. 
  The density maps and the supercluster catalogues were
calculated at the High Performance Computing Centre, University of Tartu.
In this paper we use {\it R}, an open-source free statistical environment 
developed under the GNU GPL \citep[][\texttt{http://www.r-project.org}]{ig96}. 

\end{acknowledgements}

\bibliographystyle{aa}
\bibliography{morfbib.bib}

\begin{appendix}

\section{Luminosity density field and superclusters}
\label{sec:DF}

To calculate the luminosity density field, we must  first calculate
the luminosities of groups. In flux-limited samples galaxies outside
the observational window remain unobserved, and we have also to take into account
the luminosities of these galaxies as well.  For that, we multiply the observed galaxy
luminosities by the luminosity weight $W_d$.  The distance-dependent weight
factor $W_d$ was calculated as follows:
\begin{equation}
    W_d =  {\frac{\int_0^\infty L\,n
    (L)\mathrm{d}L}{\int_{L_1}^{L_2} L\,n(L)\mathrm{d}L}} ,
    \label{eq:weight}
\end{equation}
where $L_{1,2}=L_{\sun} 10^{0.4(M_{\sun}-M_{1,2})}$ are the luminosity 
limits of the observational window at a distance $d$, corresponding to the 
absolute magnitude limits of the survey $M_1$ and $M_2$; we took 
$M_{\sun}=4.64$\,mag in the $r$-band \citep{2007AJ....133..734B}. 
Owing to their peculiar velocities, 
the distances of galaxies are somewhat uncertain; if the galaxy belongs to a 
group, we used the group distance to determine the weight factor. 

To calculate a luminosity density field, 
we converted the spatial positions of galaxies $\mathbf{r}_i$ 
and their luminosities  $L_i$ into
spatial (luminosity) densities. For that we use kernel densities
\citep{silverman86}:
\begin{equation}
    \rho(\mathbf{r}) = \sum_i K\left( \mathbf{r} - \mathbf{r}_i; a\right) L_i,
\end{equation}
where the sum is over all galaxies, and $K\left(\mathbf{r};
a\right)$ is a kernel function of a width $a$. Good kernels
for calculating densities on a spatial grid are generated by box splines
$B_J$. Box splines are local and they are interpolating on a grid:
\begin{equation}
    \sum_i B_J \left(x-i \right) = 1,
\end{equation}
for any $x$ and a small number of indices that give non-zero values for $B_J(x)$.
We used the popular $B_3$ spline function:
\begin{equation}
    B_3(x) = \frac{1}{12} \left(|x-2|^3 - 4|x-1|^3 + 6|x|^3 - 4|x+1|^3 + |x+2|^3\right).
\end{equation}
We defined the (one-dimensional) $B_3$ box spline kernel $K_B^{(1)}$ of the width $a$ as
\begin{equation}
    K_B^{(1)}(x;a,\delta) = B_3(x/a)(\delta / a),
\end{equation}
where $\delta$ is the grid step. This kernel differs from zero only
in the interval $x\in[-2a,2a]$; it is close to a Gaussian with $\sigma=1$ in the
region $x\in[-a,a]$, so its effective width is $2a$ \citep[see, e.g.,][]{saar09}.
The kernel exactly preserves the
interpolation property for all values of $a$ and $\delta$,
where the ratio $a/\delta$ is an integer. (This kernel can be used also if this ratio
is not an integer, 
and $a \gg \delta$; the kernel sums to 1 in this case, too, with a very small error).
This means that if we apply this kernel to $N$ points on a one-dimensional grid,
the sum of the densities over the grid is exactly $N$.
 
The three-dimensional kernel $K_B^{(3)}$
is given by the direct product of three one-dimensional kernels:
\begin{equation}
    K_B^{(3)}(\mathbf{r};a,\delta) \equiv K_3^{(1)}(x;a,\delta) K_3^{(1)}(y;a,\delta) K_3^{(1)}(z;a,\delta),
\end{equation}
where $\mathbf{r} \equiv \{x,y,z\}$. Although this is a direct product,
it is isotropic to a good degree \citep{saar09}.

The densities were calculated on a cartesian grid based on the SDSS $\eta$,
$\lambda$ coordinate system, because it allowed the most efficient fit of
the galaxy sample cone into a brick.
Using the rms velocity $\sigma_v$, translated into distance,
and the rms projected radius $\sigma_r$ from the group catalogue (T10),
we suppressed the cluster finger redshift distortions. We
divided the radial distances between the group galaxies and the group centre by the ratio
of the rms sizes of the group finger:
\begin{equation}
    d_\mathrm{gal,f} = d_\mathrm{group} + (d_\mathrm{gal,i} - 
    d_\mathrm{group})\; \sigma_\mathrm{r} / \sigma_\mathrm{v}.
\end{equation}
This removes the smudging effect the fingers have on the density field.
We used 
an 1~\Mpc\ step grid and chose the kernel width $a=8$~\Mpc.
This kernel differs from zero within the radius 16~\Mpc,
but significantly so only inside the 8~\Mpc\ radius.

Before extracting superclusters we applied the DR7 mask 
constructed by P.~Arnalte-Mur \citep{martinez09, 
2010arXiv1012.1989J} to the density field
and converted densities into units of mean density. The mean
density is defined as the average over all pixel values inside the mask. The mask is
designed to follow the edges of the survey and the galaxy distribution inside the
mask is assumed to be homogeneous.

\section{Minkowski functionals and shapefinders} 
\label{sec:MF}

For a given surface the four Minkowski functionals (from the first to the
fourth) are proportional to the enclosed volume $V$, the area of the surface
$S$, the integrated mean curvature $C$, and the integrated Gaussian curvature
$\chi$. 
Consider an
excursion set $F_{\phi_0}$ of a field $\phi(\mathbf{x})$ (the set
of all points where the density is higher than a given limit,
$\phi(\mathbf{x}\ge\phi_0$)). Then, the first
Minkowski functional (the volume functional) is the volume of 
this region (the excursion set):
\begin{equation}
\label{mf0}
V_0(\phi_0)=\int_{F_{\phi_0}}\mathrm{d}^3x\;.
\end{equation}
The second Minkowski functional is proportional to the surface area
of the boundary $\delta F_\phi$ of the excursion set:
\begin{equation}
\label{mf1}
V_1(\phi_0)=\frac16\int_{\delta F_{\phi_0}}\mathrm{d}S(\mathbf{x})\;,
\end{equation}
(but it is not the area itself, notice the constant).
The third Minkowski functional is proportional to the
integrated mean curvature 
$C$ of the boundary:
\begin{equation}
\label{mf2}
V_2(\phi_0)=\frac1{6\pi}\int_{\delta F_{\phi_0}}
    \left(\frac1{R_1(\mathbf{x})}+\frac1{R_2(\mathbf{x})}\right)\mathrm{d}S(\mathbf{x})\;,
\end{equation}
where $R_1(\mathbf{x})$ and $R_2(\mathbf{x})$ 
are the principal radii of curvature of the boundary.
The fourth Minkowski functional is proportional to the integrated
Gaussian curvature (the Euler characteristic) 
of the boundary:
\begin{equation}
\label{mf3}
V_3(\phi_0)=\frac1{4\pi}\int_{\delta F_{\phi_0}}
    \frac1{R_1(\mathbf{x})R_2(\mathbf{x})}\mathrm{d}S(\mathbf{x})\;.
\end{equation}
At high (low) densities this functional gives us the number of isolated 
clumps (void bubbles) in the sample 
\citep{mar03,saar06}:

\begin{equation}
\label{v3}
V_3=N_{\mbox{clumps}} + N_{\mbox{bubbles}} - N_{\mbox{tunnels}}.
\end{equation}

As the argument labelling the isodensity surfaces, we chose the (excluded) mass
fraction $mf$ -- the ratio of the mass in the regions with the density {\em lower}
than the density at the surface, to the total mass of the supercluster. When
this ratio runs from 0 to 1, the iso-surfaces move from the outer limiting
boundary into the centre of the supercluster, i.e., the fraction $mf=0$
corresponds to the whole supercluster, and $mf=1$ -- to its highest density
peak.

We used directly only the fourth Minkowski functional in this paper;
the other functionals were used to calculate the shapefinders
\citep{sah98,sss04,saar09}.  The shapefinders are defined as a set of
combinations of Minkowski functionals: $H_1=3V/S$ (thickness),
$H_2=S/C$ (width), and $H_3=C/4\pi$ (length).  The shapefinders have
dimensions of length and are normalized to give $H_i=R$ for a sphere
of radius $R$.  For 
smooth (ellipsoidal) surfaces, the shapefinders $H_i$
follow the inequalities $H_1\leq H_2\leq H_3$.  Oblate ellipsoids (pancakes)
are characterised by $H_1 << H_2 \approx H_3$, while prolate ellipsoids
(filaments) are described by $H_1 \approx H_2 << H_3$.

\citet{sah98} also defined  two dimensionless
shapefinders $K_1$ (planarity) and $K_2$ (filamentarity): 
$K_1 = (H_2 - H_1)/(H_2 + H_1)$ and $K_2 = (H_3 -
H_2)/(H_3 + H_2)$.

In the $(K_1,K_2)$-plane filaments are located near the $K_2$-axis,
pancakes near the $K_1$-axis, and ribbons along the diameter, connecting 
the spheres at the origin with the ideal ribbon at $(1,1)$. 
In \citet{e07} we calculated typical morphological signatures 
of a series of empirical models that serve as  morphological 
templates to compare with the characteristic curves for superclusters
in the $(K_1,K_2)$-plane.

\section{3D figures of superclusters}
\label{sec:3dfig}

\begin{figure}[ht]
\centering
\resizebox{0.42\textwidth}{!}{\includegraphics*{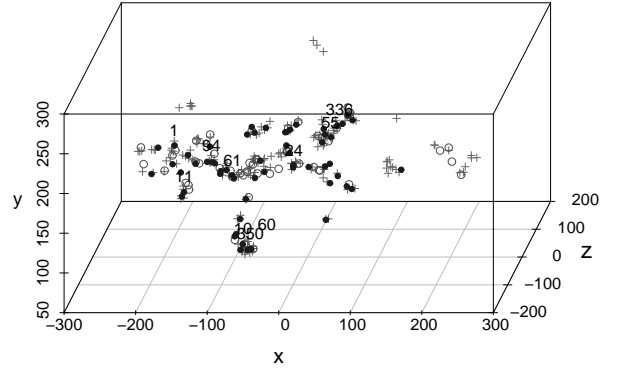}}
\caption{Distribution of groups with at least 12 member galaxies
in our superclusters in the cartesian coordinates as in
Fig.~\ref{fig:sclid}, in units of \Mpc. 
Filled circles denote the groups with at least 50 member galaxies,
empty circles denote the groups with 30--49 member galaxies
and crosses denote the groups with 12--29 member galaxies. 
The numbers show the ID's (Table~\ref{tab:scldata}, column 1)
of very rich superclusters from Sect.~\ref{sect:sclnotes}.
}
\label{fig:3dall}
\end{figure}

\begin{figure}[ht]
\centering
\resizebox{0.22\textwidth}{!}{\includegraphics*{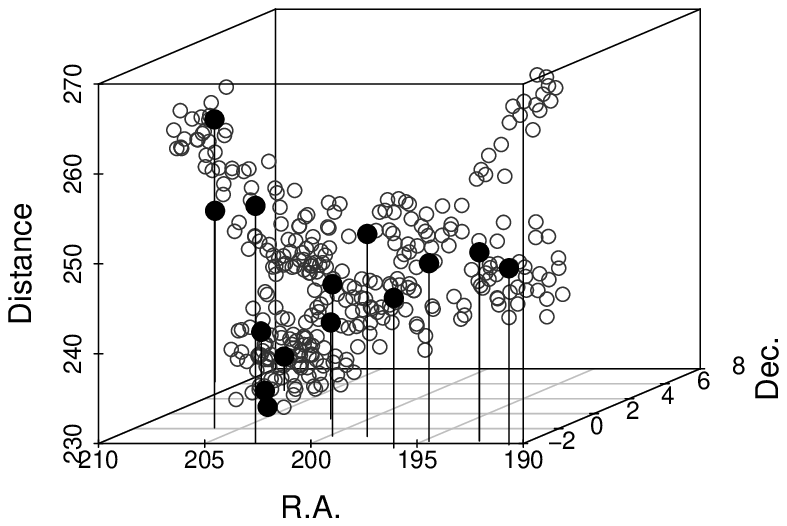}}
\resizebox{0.22\textwidth}{!}{\includegraphics*{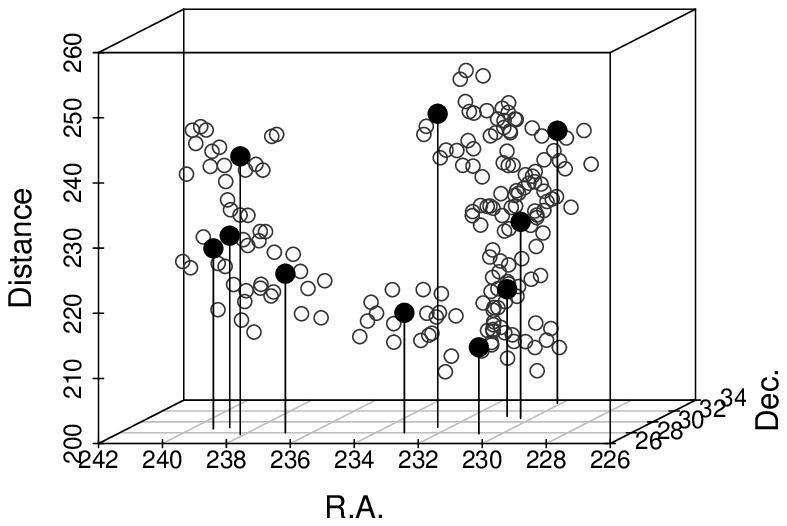}}\\
\resizebox{0.22\textwidth}{!}{\includegraphics*{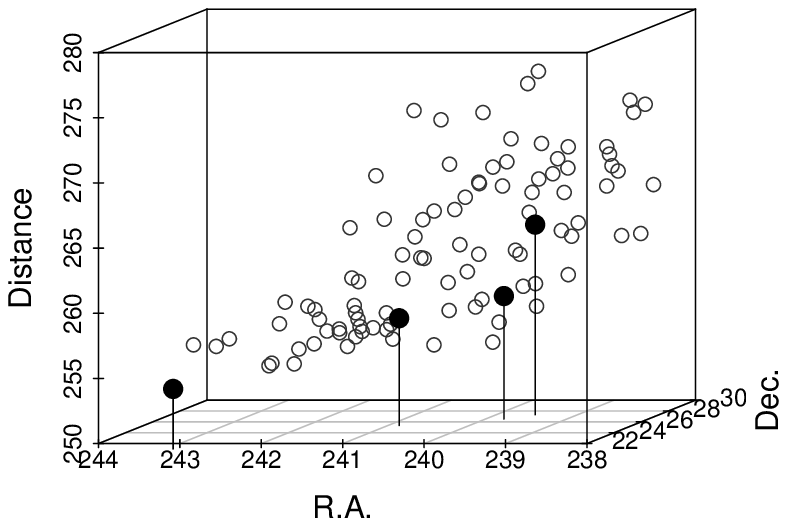}}
\resizebox{0.22\textwidth}{!}{\includegraphics*{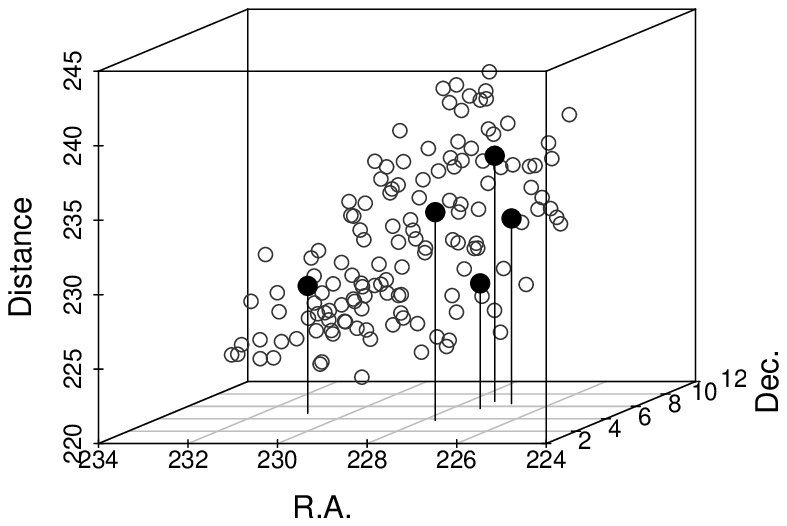}}\\
\resizebox{0.22\textwidth}{!}{\includegraphics*{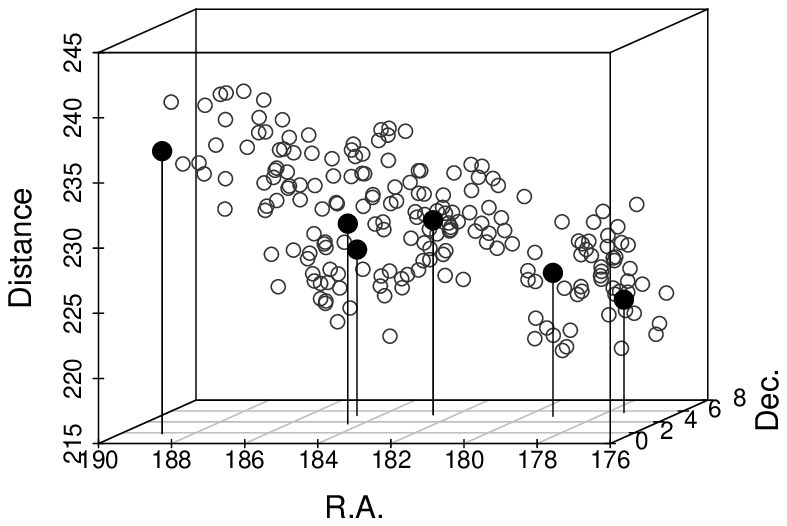}}
\resizebox{0.22\textwidth}{!}{\includegraphics*{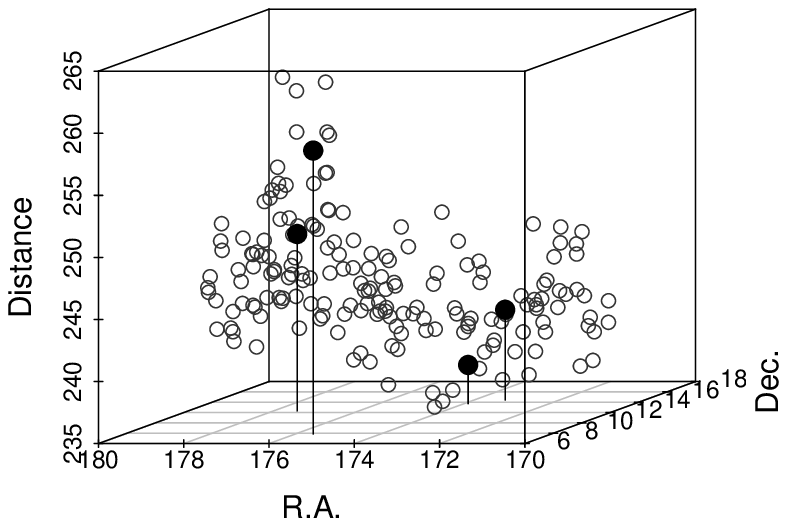}}\\
\resizebox{0.22\textwidth}{!}{\includegraphics*{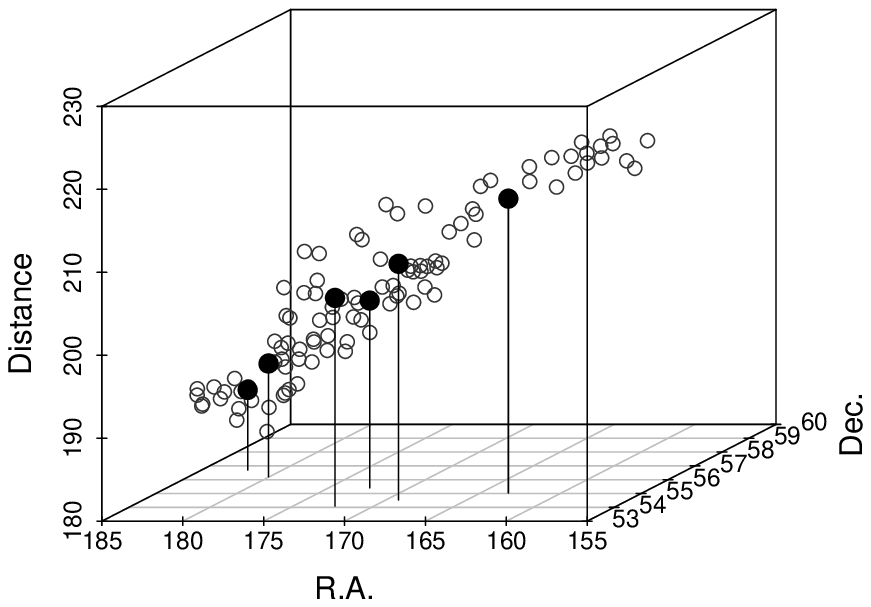}}
\resizebox{0.22\textwidth}{!}{\includegraphics*{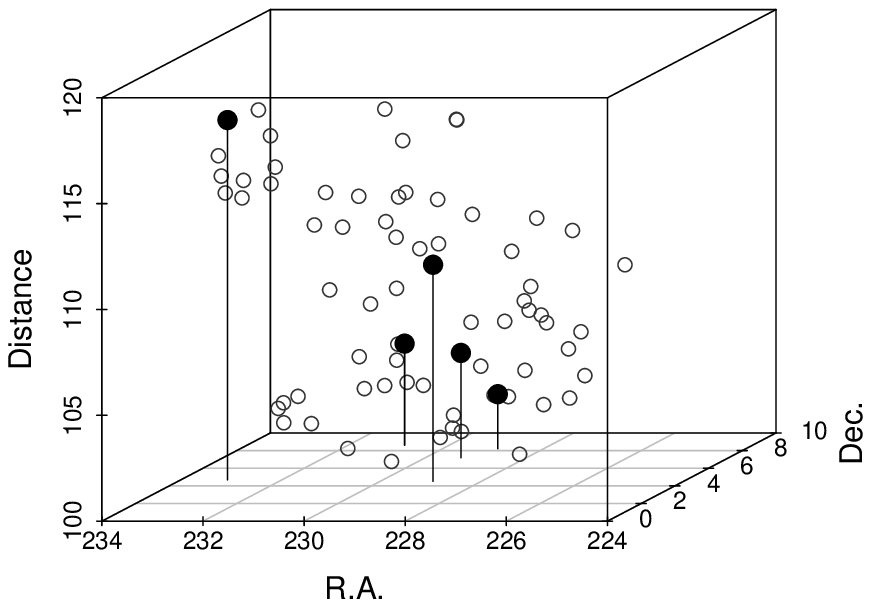}}\\
\caption{3D figures of the very rich superclusters described in
Sect.~\ref{sect:sclnotes}. Filled circles show the location of rich groups and
clusters with at least 30 member galaxies, empty circles show poorer groups.
We plot the right ascension (in degrees), declination (in degrees),
and distance (in Mpc/$h$) of groups. 
Plots of superclusters are given in the same order as they are 
presented in the text. From top to bottom:
left: the supercluster SCl~061, right: SCl~094, 
left: the supercluster SCl~001, right: SCl~011, 
left: the supercluster SCl~024, right: SCl~055, 
left: the supercluster SCl~336, right: SCl~350. 
}
\label{fig:3d1}
\end{figure}

\begin{figure}[ht]
\centering
\resizebox{0.22\textwidth}{!}{\includegraphics*{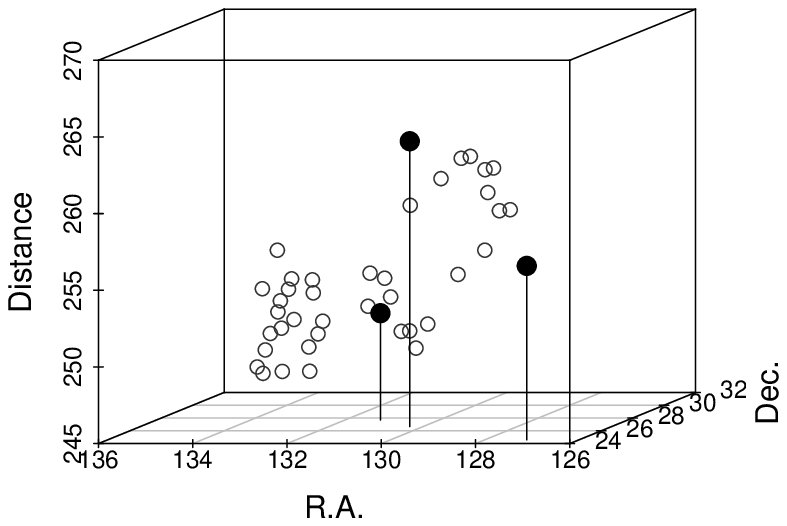}}
\resizebox{0.22\textwidth}{!}{\includegraphics*{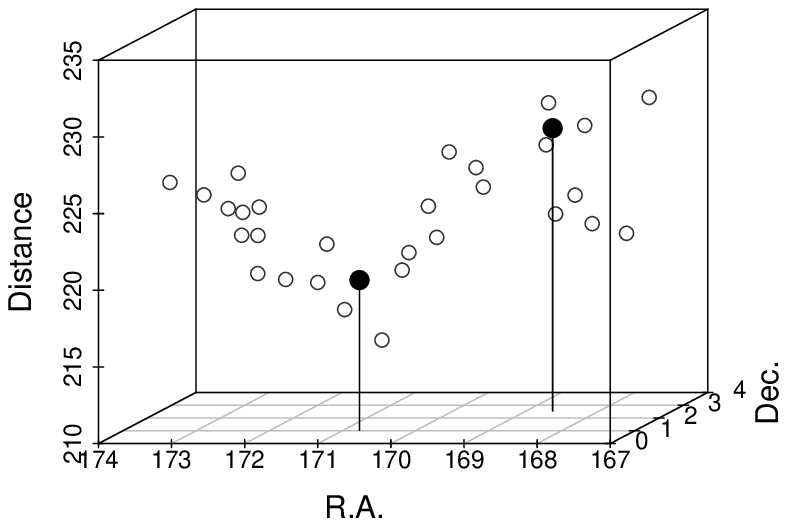}}\\
\resizebox{0.22\textwidth}{!}{\includegraphics*{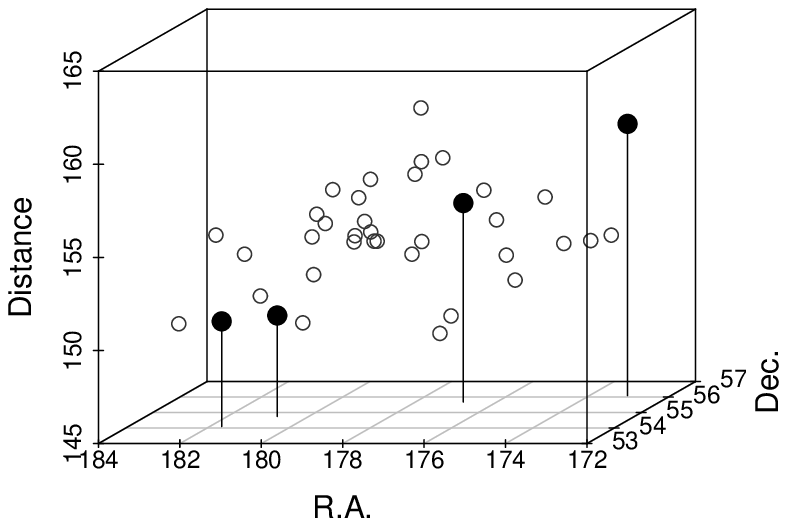}}
\resizebox{0.22\textwidth}{!}{\includegraphics*{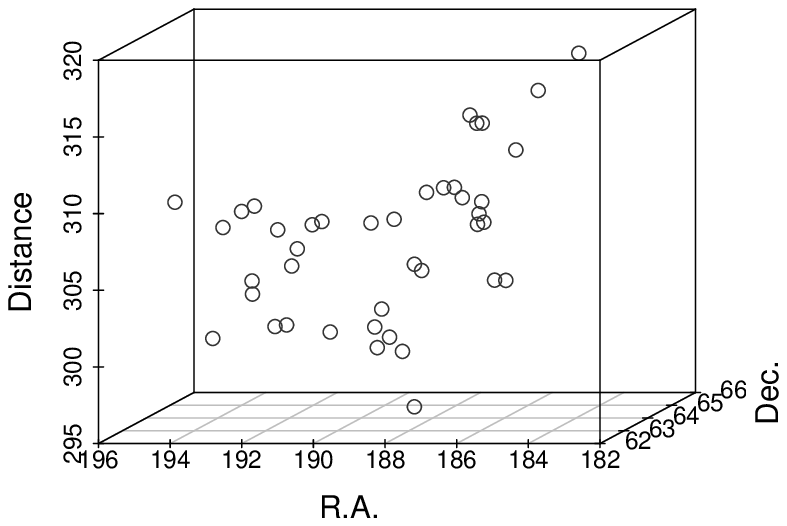}}\\
\resizebox{0.22\textwidth}{!}{\includegraphics*{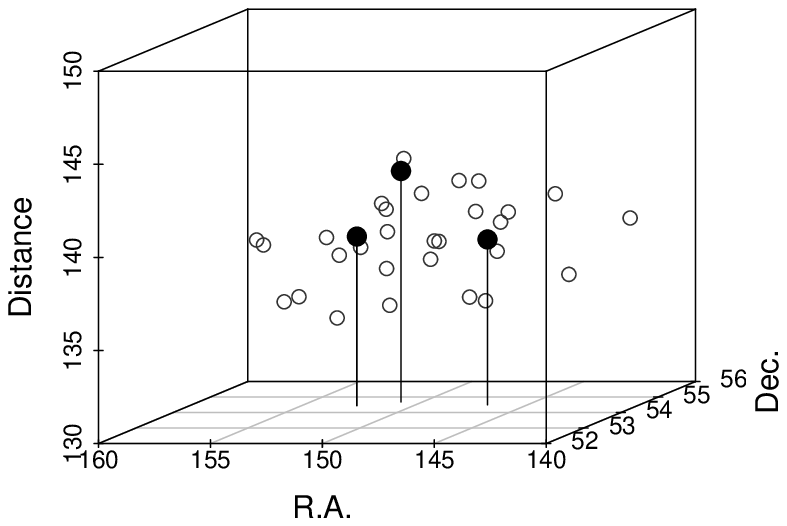}}
\resizebox{0.22\textwidth}{!}{\includegraphics*{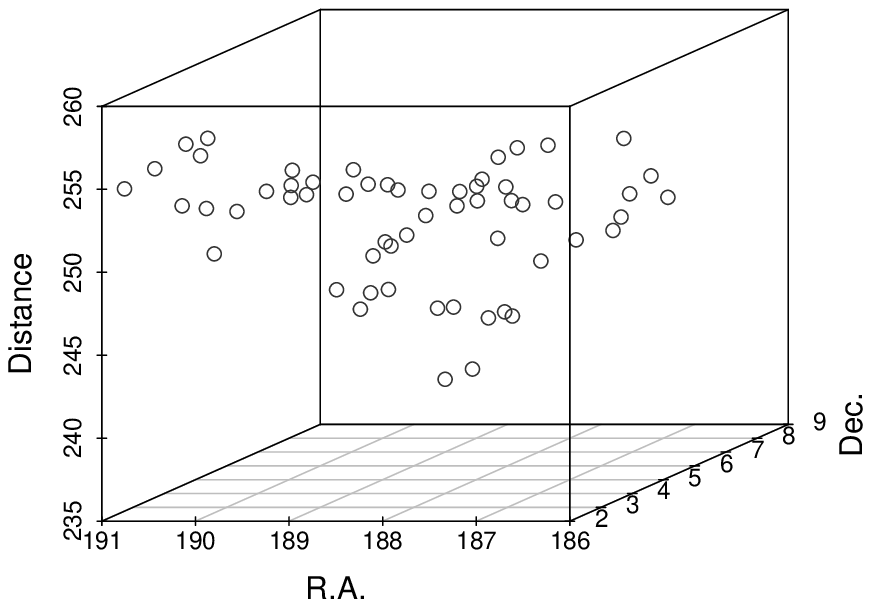}}\\
\caption{3D figures of superclusters with less than 950 member galaxies (continued).  
Panels and notations in this figure  are the same as in Fig.~\ref{fig:3d1}. 
From top to bottom:
left: the supercluster SCl~474, right: SCl~512, 
left: the supercluster SCl~525, right: SCl~530, 
left: the supercluster SCl~779, right: SCl~827. 
}
\label{fig:3d3}
\end{figure}

Here we  show the 3D distribution of groups in our superclusters. 
Figure~\ref{fig:3dall} is 3D presentation of Fig.~\ref{fig:sclid}
and shows the distribution of groups with at least twelve member galaxies
in our superclusters in the cartesian coordinates defined in Section~\ref{sect:LSS}, 
in units of \Mpc. 
The filled circles denote groups with at least 50 member galaxies,
empty circles denote groups with 30--49 member galaxies
and crosses denote groups with 12--29 member galaxies.
The numbers are 
ID numbers of superclusters with at least 950 member galaxies.
Fig.~\ref{fig:3d1}, ~\ref{fig:3d3}, and ~\ref{fig:3d3}  show
3D distribution of groups in superclusters. In these figures
we plot rich groups with
at least 30 member galaxies with filled circles, 
groups with 2 -- 29 member  galaxies with empty circles. We do not
show the figures for superclusters, for which
the value of the fourth Minkowski functional $V_3 = 1$ 
over the whole mass fraction interval. 

\begin{figure}[ht]
\centering
\resizebox{0.22\textwidth}{!}{\includegraphics*{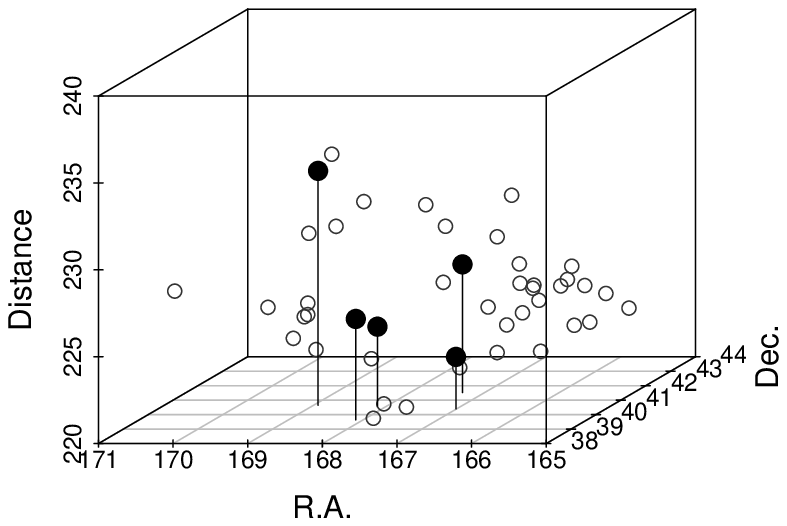}}
\resizebox{0.22\textwidth}{!}{\includegraphics*{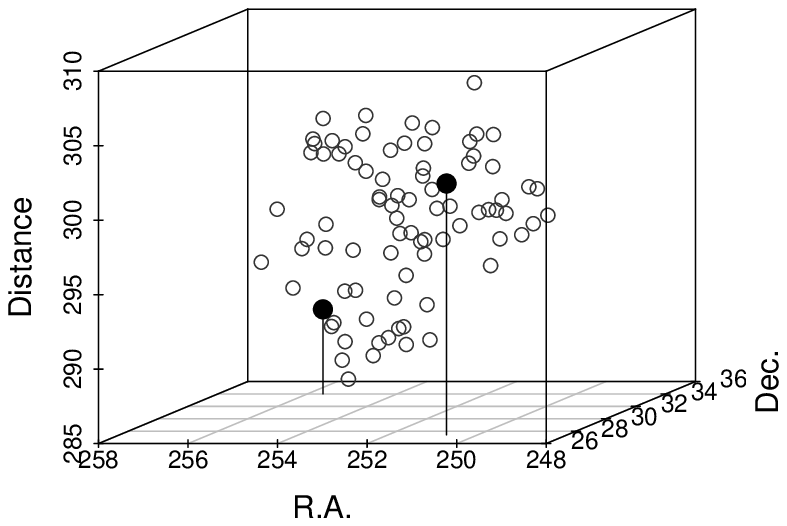}}\\
\resizebox{0.22\textwidth}{!}{\includegraphics*{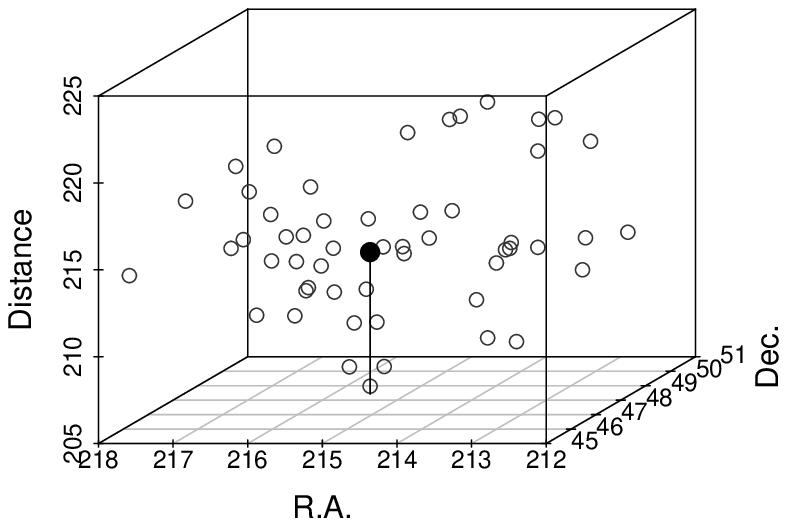}}
\resizebox{0.22\textwidth}{!}{\includegraphics*{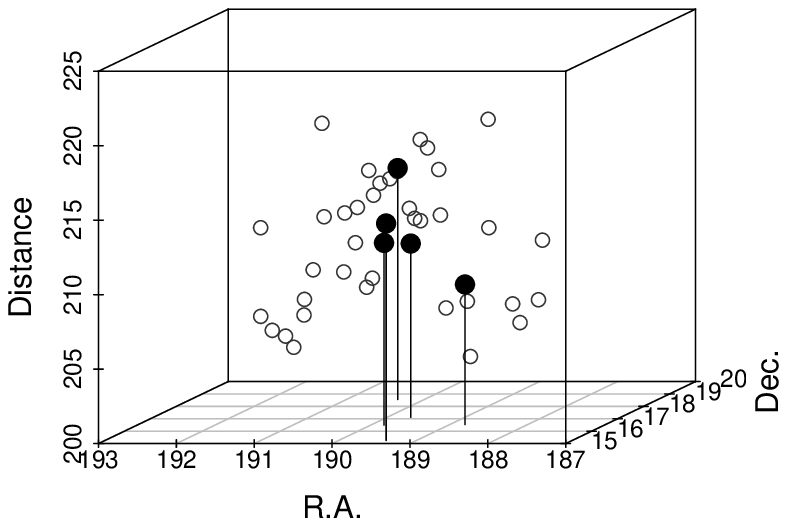}}\\
\resizebox{0.22\textwidth}{!}{\includegraphics*{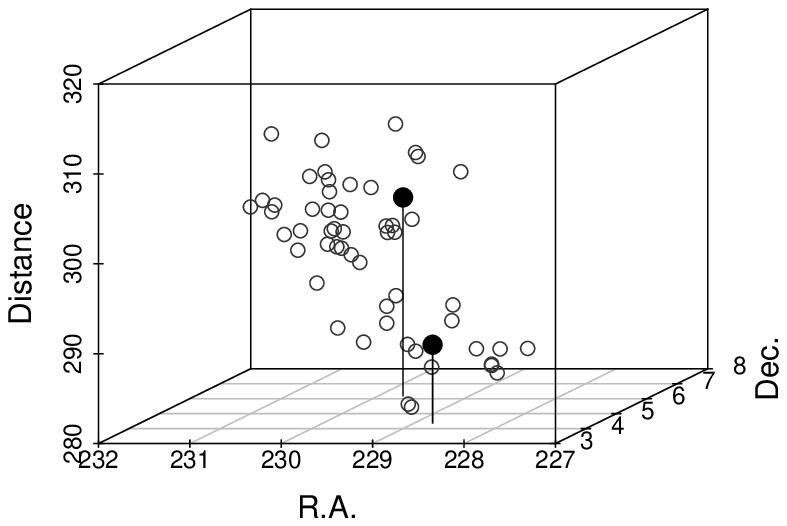}}
\resizebox{0.22\textwidth}{!}{\includegraphics*{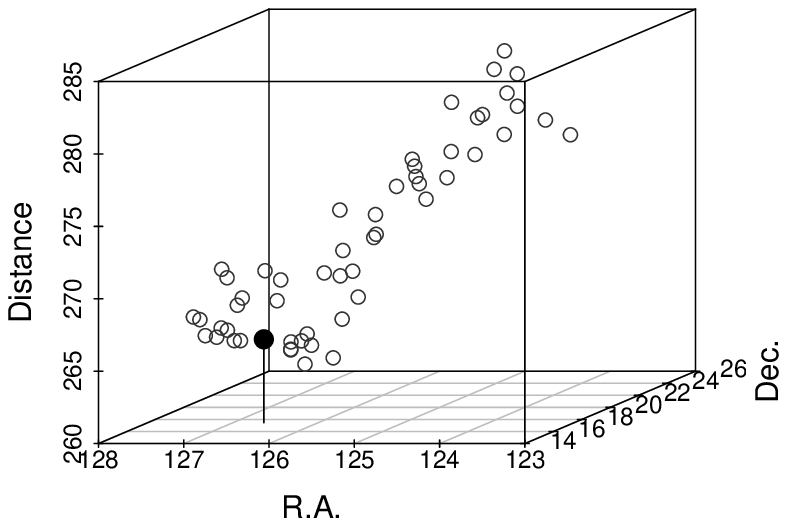}}\\
\resizebox{0.22\textwidth}{!}{\includegraphics*{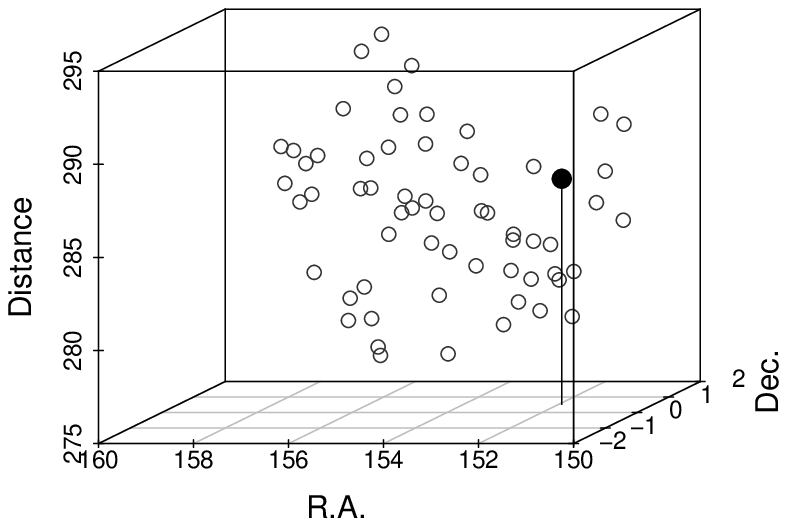}}
\resizebox{0.22\textwidth}{!}{\includegraphics*{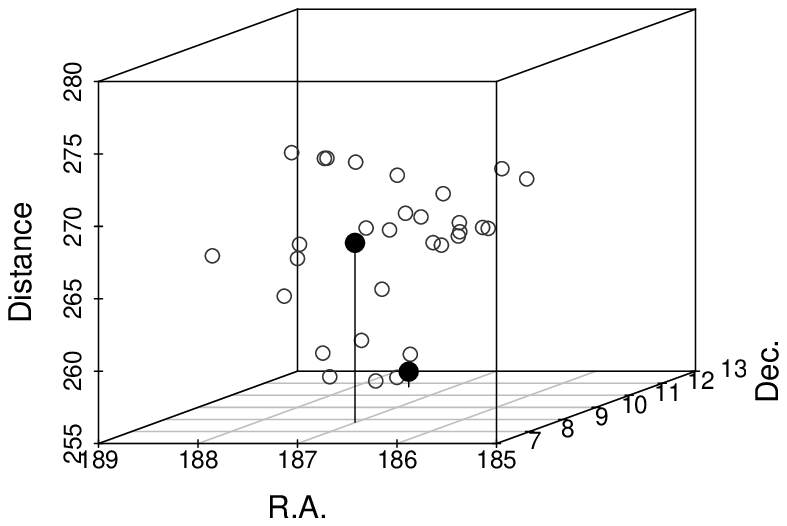}}\\
\resizebox{0.22\textwidth}{!}{\includegraphics*{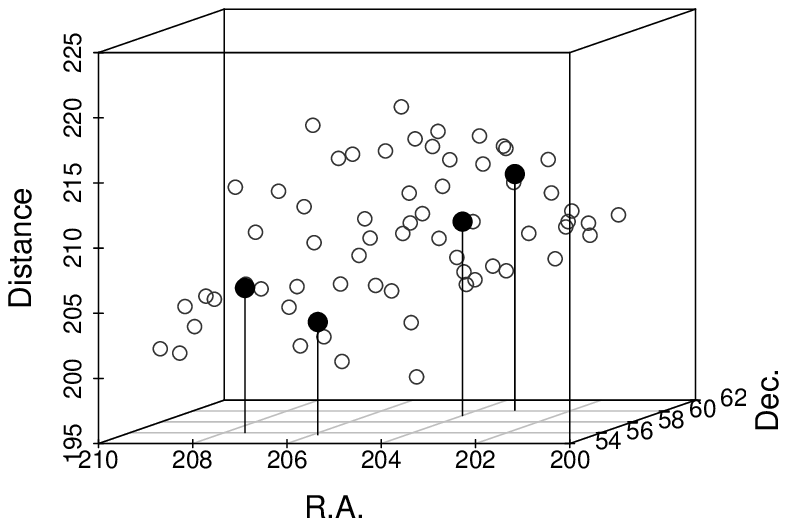}}
\resizebox{0.22\textwidth}{!}{\includegraphics*{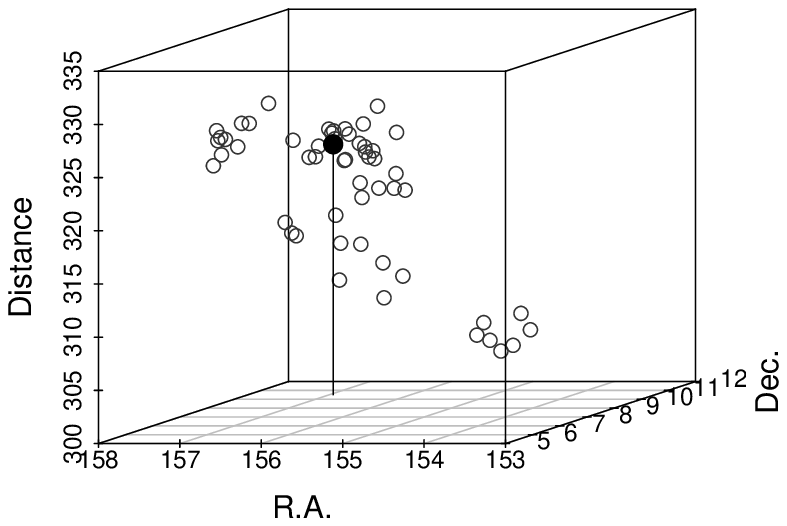}}\\
\resizebox{0.22\textwidth}{!}{\includegraphics*{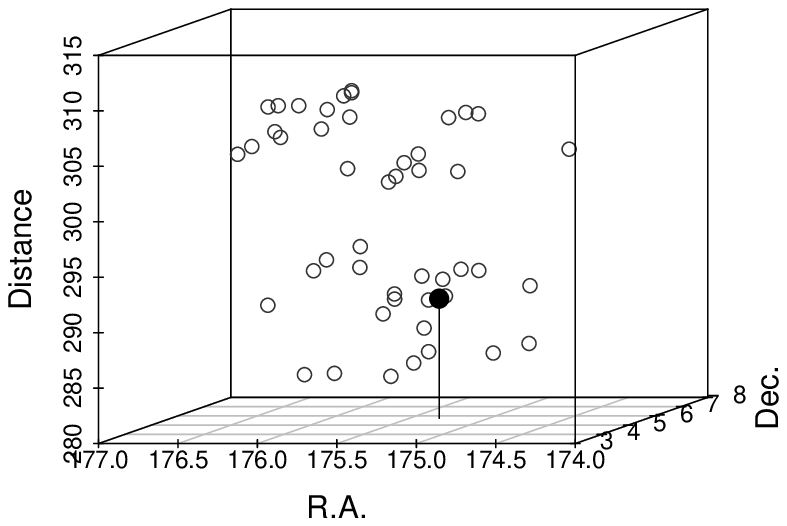}}
\resizebox{0.22\textwidth}{!}{\includegraphics*{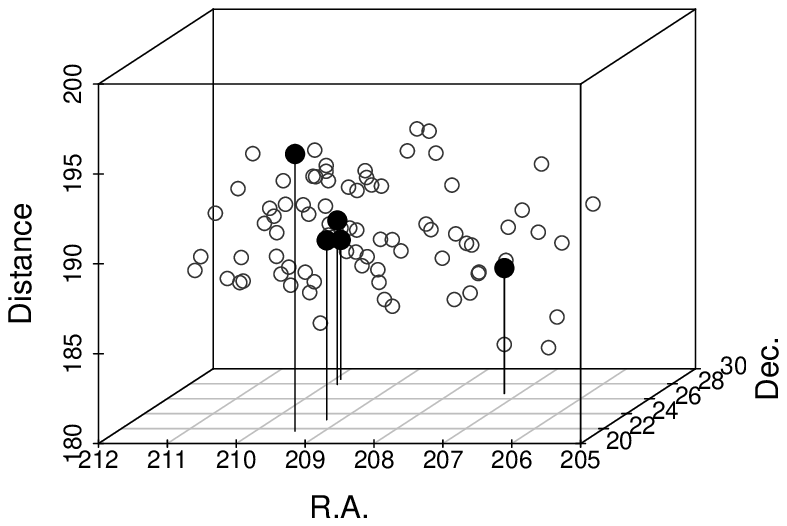}}\\
\resizebox{0.22\textwidth}{!}{\includegraphics*{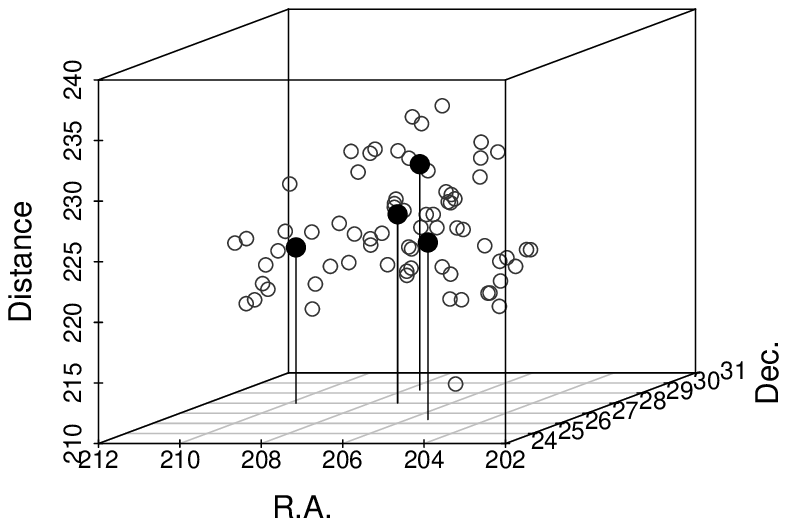}}
\resizebox{0.22\textwidth}{!}{\includegraphics*{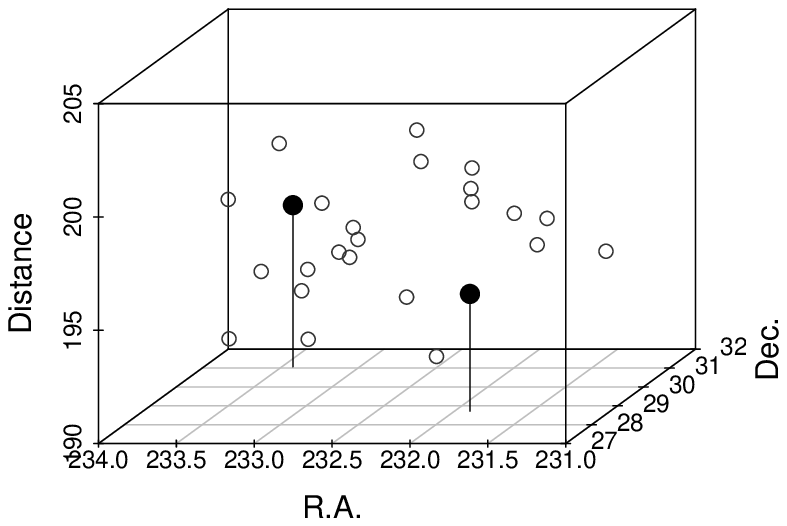}}\\
\resizebox{0.22\textwidth}{!}{\includegraphics*{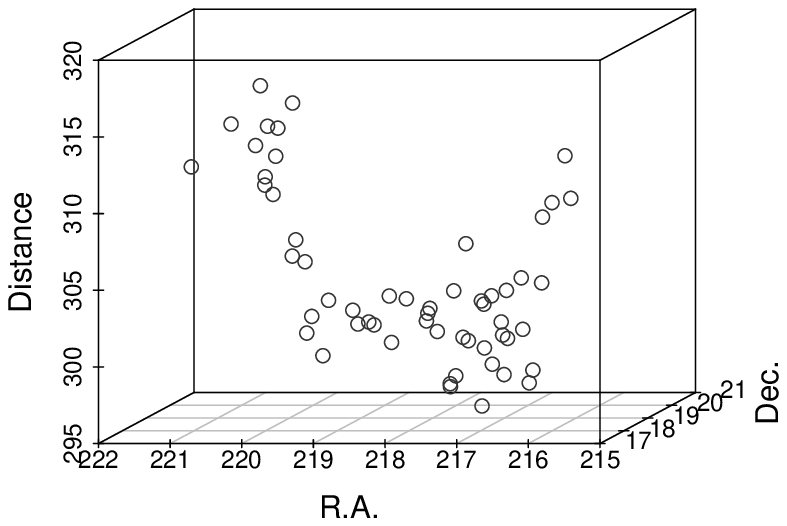}}
\resizebox{0.22\textwidth}{!}{\includegraphics*{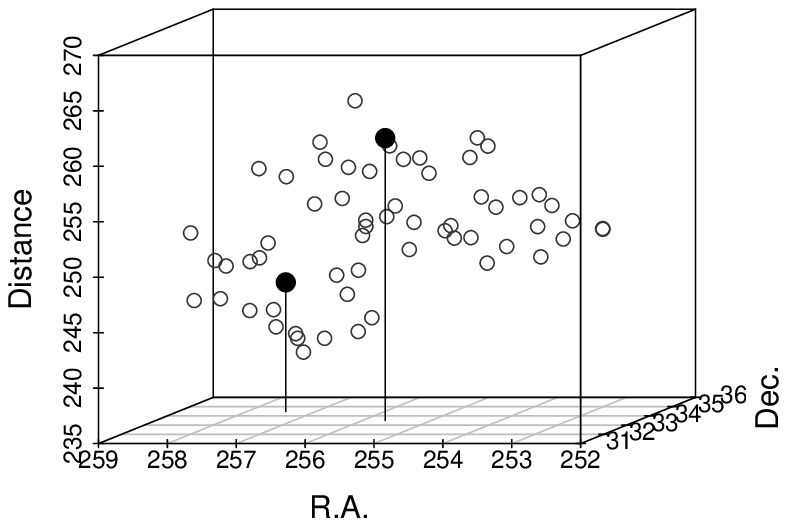}}\\
\caption{3D figures of superclusters with less than 950 member galaxies. 
Panels and notations in this figure  are the same as in Fig.~\ref{fig:3d1}. 
From top to bottom:
left: the supercluster SCl~038, right: SCl~064, 
left: the supercluster SCl~087, right: SCl~136, 
left: the supercluster SCl~152, right: SCl~189, 
left: the supercluster SCl~198, right: SCl~223, 
left: the supercluster SCl~228, right: SCl~317, 
left: the supercluster SCl~332, right: SCl~349, 
left: the supercluster SCl~351, right: SCl~362, 
left: the supercluster SCl~366, right: SCl~376. 
}
\label{fig:3d2}
\end{figure}

\section{The distribution of galaxies and rich groups in the sky and
the fourth Minkowski functional and morphological signature for 
superclusters with less than 950 member galaxies} 
\label{sec:MFfig}

\begin{figure}[ht]
\centering
\resizebox{0.42\textwidth}{!}{\includegraphics*{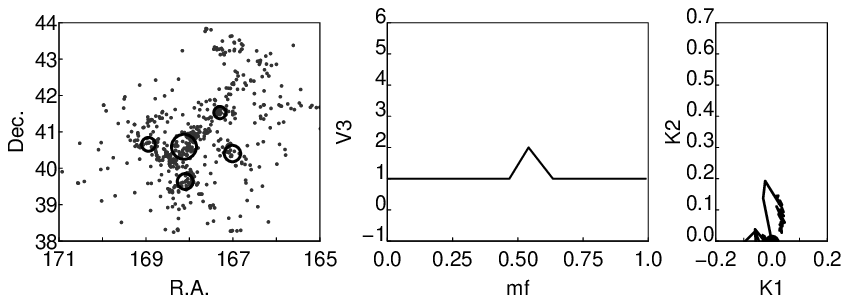}}\\
\resizebox{0.42\textwidth}{!}{\includegraphics*{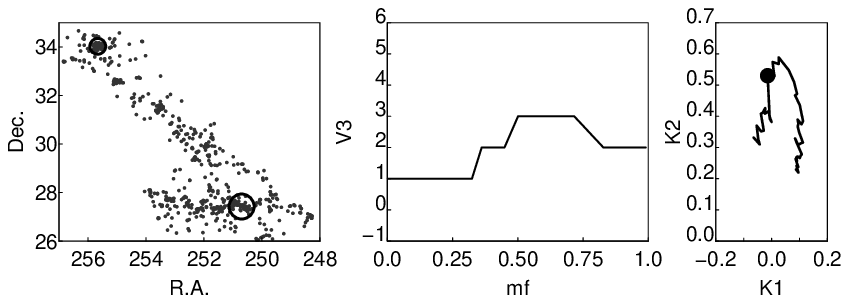}}\\
\resizebox{0.42\textwidth}{!}{\includegraphics*{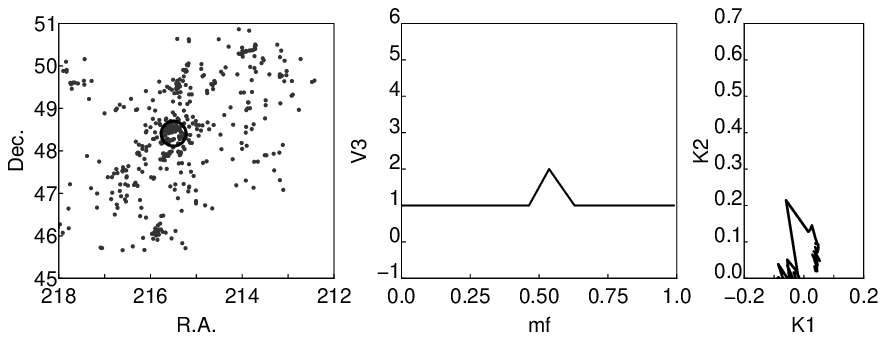}}\\
\resizebox{0.42\textwidth}{!}{\includegraphics*{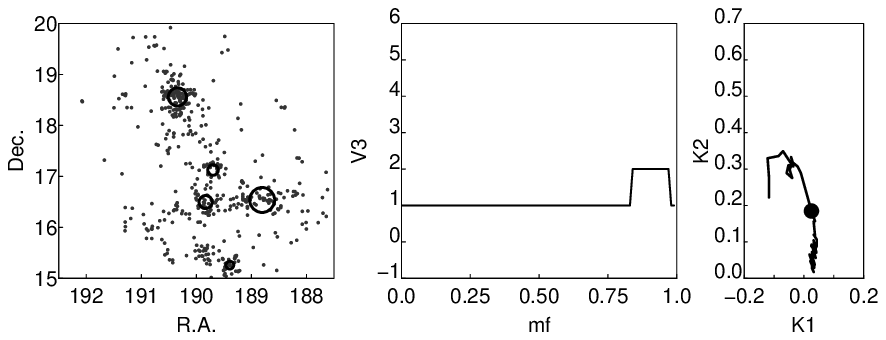}}\\
\resizebox{0.42\textwidth}{!}{\includegraphics*{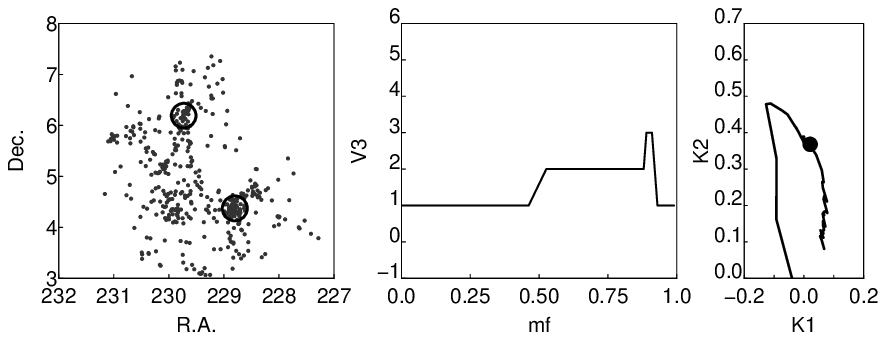}}\\
\resizebox{0.42\textwidth}{!}{\includegraphics*{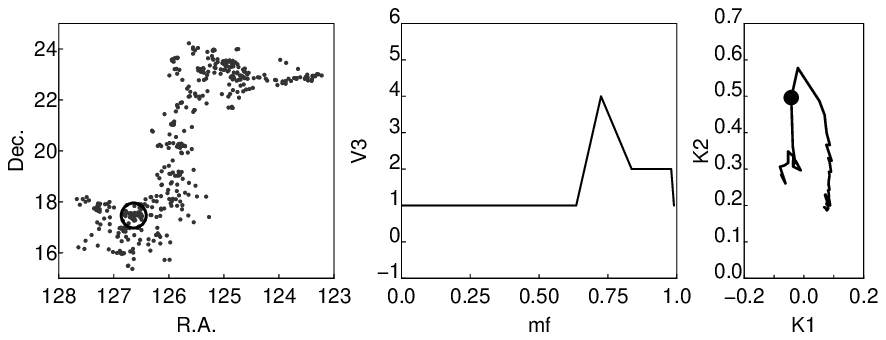}}\\
\resizebox{0.42\textwidth}{!}{\includegraphics*{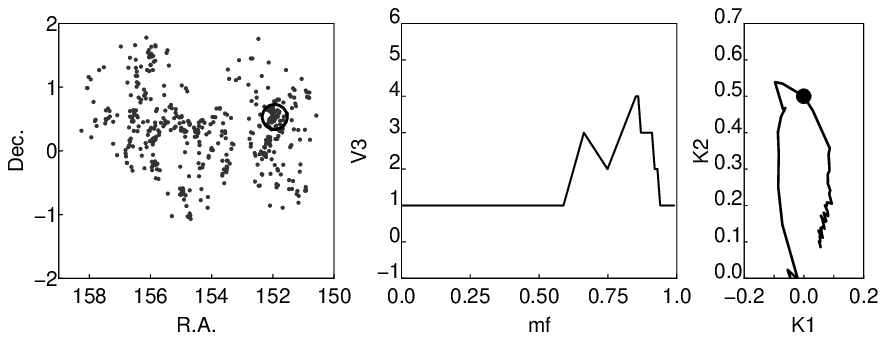}}\\
\resizebox{0.42\textwidth}{!}{\includegraphics*{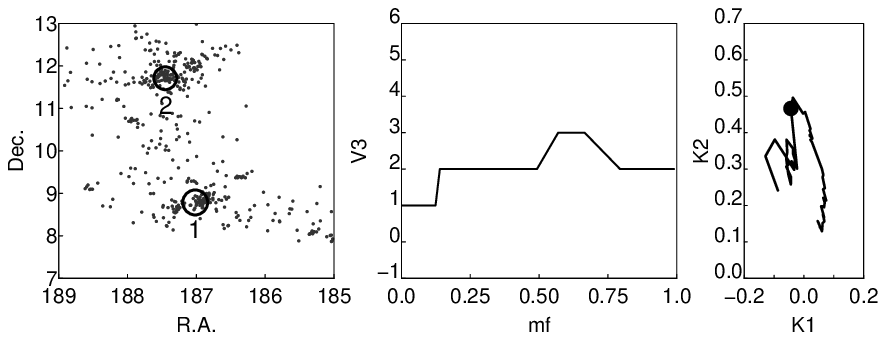}}\\
\caption{Left panels show the distribution of galaxies (dots) in the sky. 
Circles mark the location of groups with at least 
30 member galaxies, and the size of a circle is proportional to the size of a group
in the sky. 
The middle panels show the fourth Minkowski functional $V_3$ vs. the mass fraction
$mf$, and the right panels show
the shapefinders $K_1$ (planarity) and $K_2$ (filamentarity) 
(the morphological signature).
In the right panels filled circles  mark the value of the mass fraction 
$mf= 0.7$; the mass fraction increases 
anti-clockwise along the curves.
From top to bottom:
the superclusters SCl~038, SCl~064, SCl~087, SCl~136, SCl~152,  SCl~189,
SCl~198, and SCl~223. 
}
\label{fig:sclapp1}
\end{figure}

\begin{figure}[ht]
\centering
\resizebox{0.42\textwidth}{!}{\includegraphics*{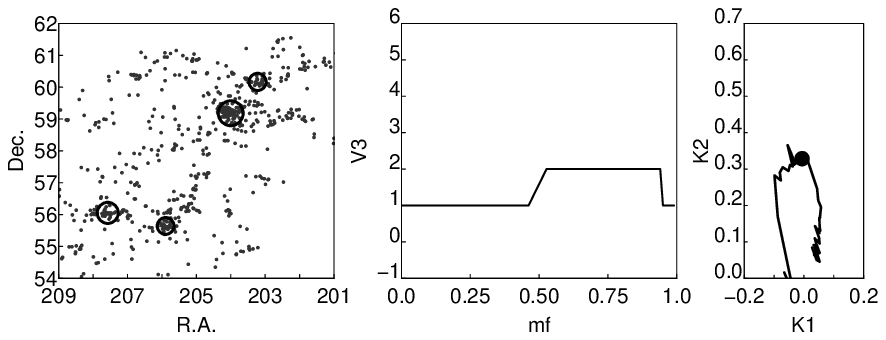}}\\
\resizebox{0.42\textwidth}{!}{\includegraphics*{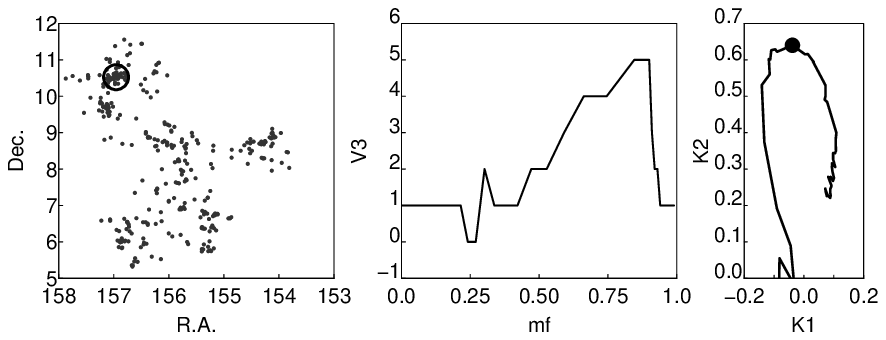}}\\
\resizebox{0.42\textwidth}{!}{\includegraphics*{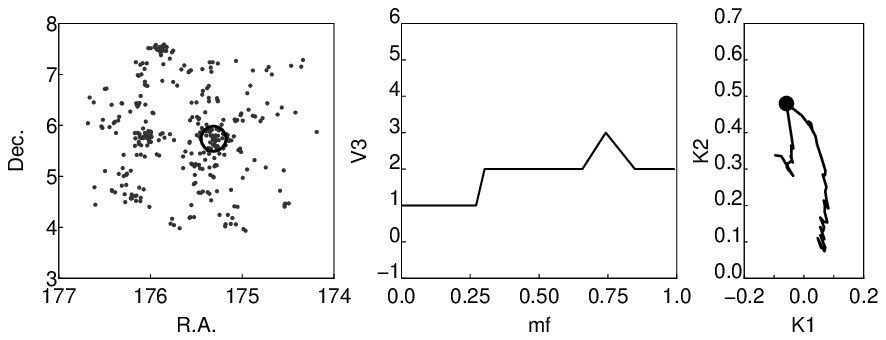}}\\
\resizebox{0.42\textwidth}{!}{\includegraphics*{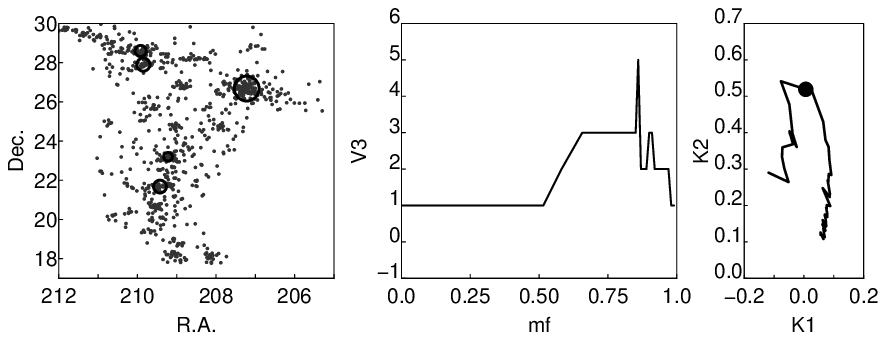}}\\
\resizebox{0.42\textwidth}{!}{\includegraphics*{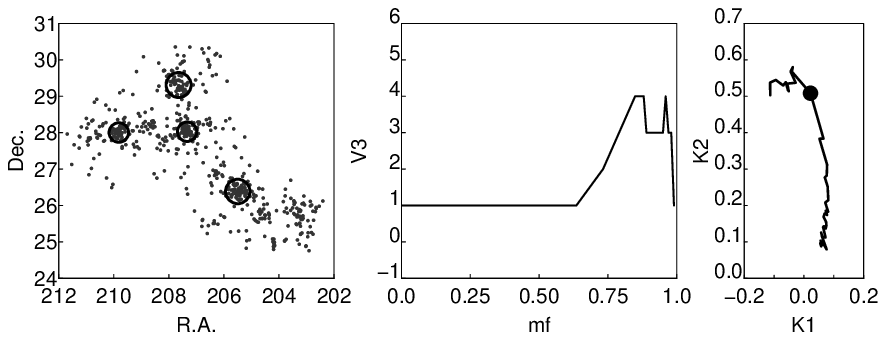}}\\
\resizebox{0.42\textwidth}{!}{\includegraphics*{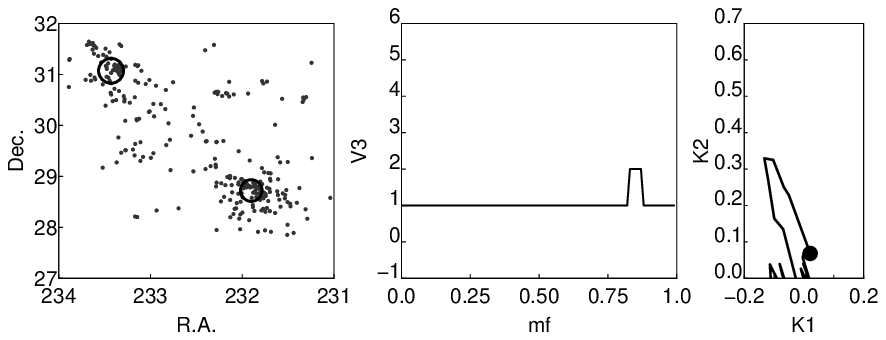}}\\
\resizebox{0.42\textwidth}{!}{\includegraphics*{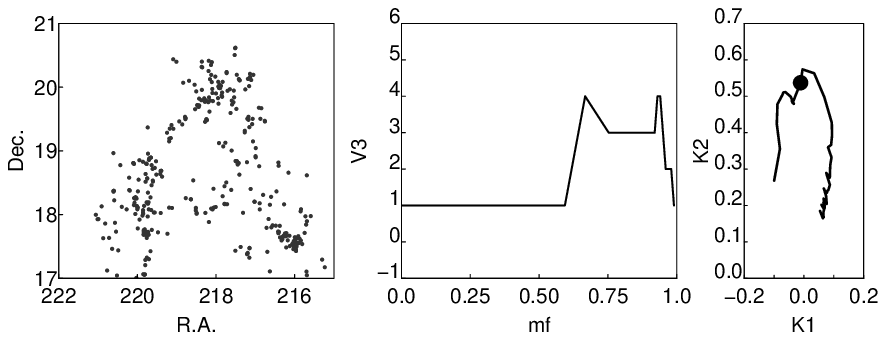}}\\
\resizebox{0.42\textwidth}{!}{\includegraphics*{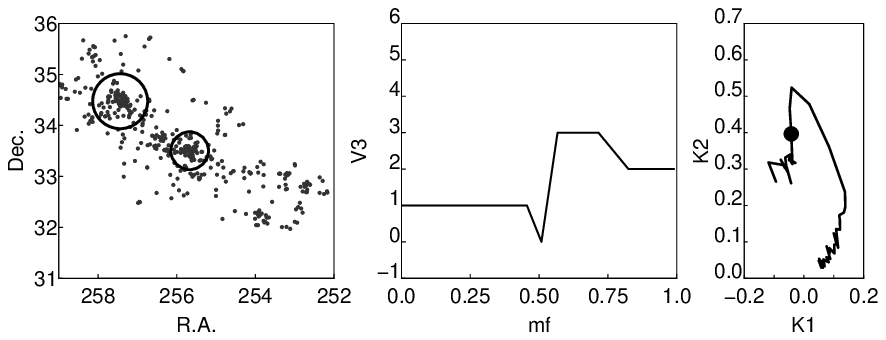}}\\
\caption{Panels in this figure  are the same as in Fig.~\ref{fig:sclapp1}.
From top to bottom:
the superclusters SCl~228, SCl~317, SCl~332, SCl~349, SCl~351,
SCl~362, SCl~366, and SCl~376. 
}
\label{fig:sclapp2}
\end{figure}

\begin{figure}[ht]
\centering
\resizebox{0.42\textwidth}{!}{\includegraphics*{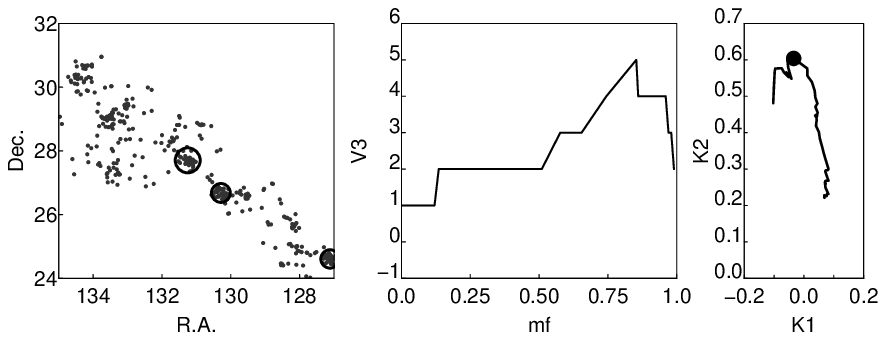}}\\
\resizebox{0.42\textwidth}{!}{\includegraphics*{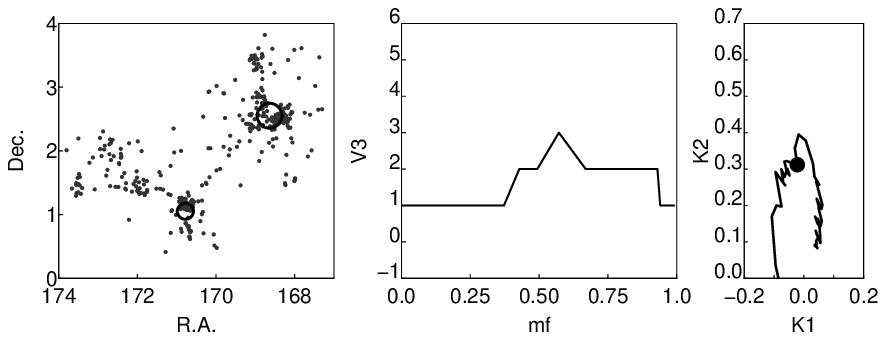}}\\
\resizebox{0.42\textwidth}{!}{\includegraphics*{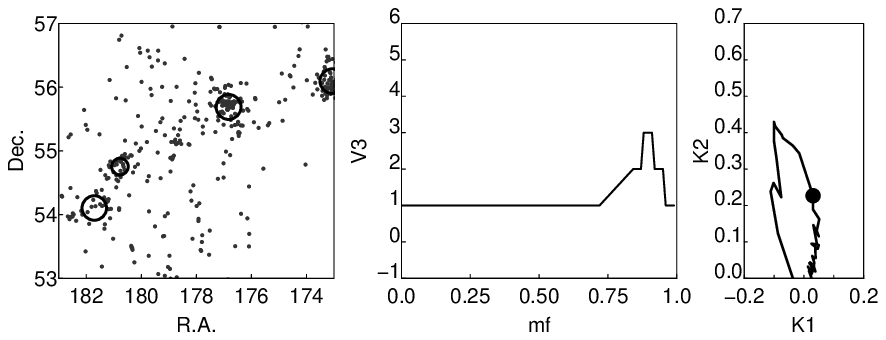}}\\
\resizebox{0.42\textwidth}{!}{\includegraphics*{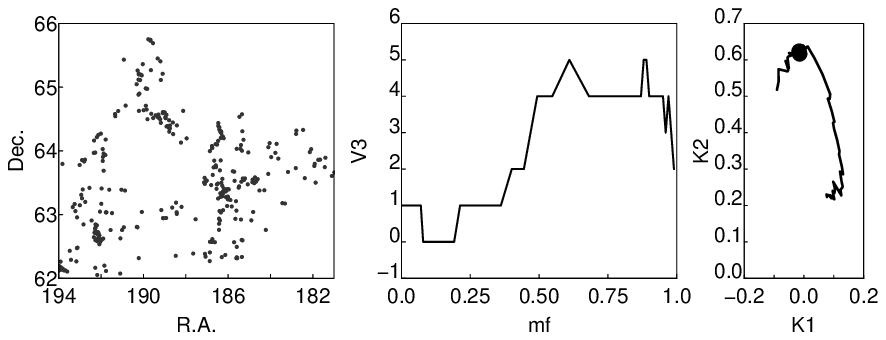}}\\
\resizebox{0.42\textwidth}{!}{\includegraphics*{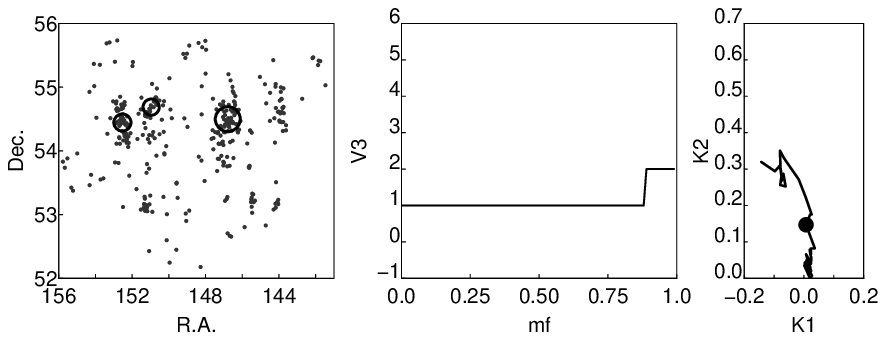}}\\
\resizebox{0.42\textwidth}{!}{\includegraphics*{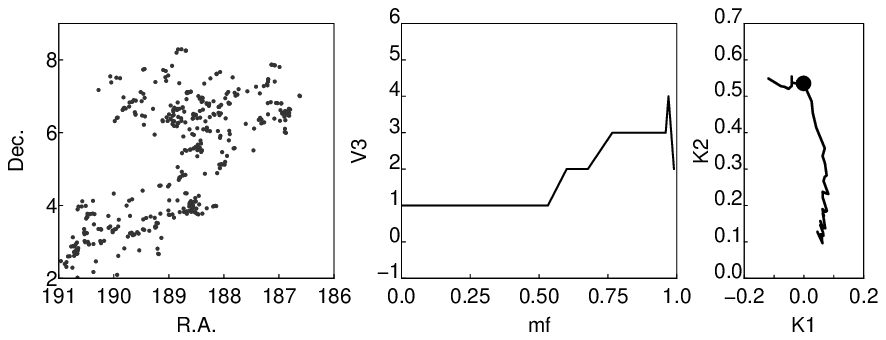}}\\
\caption{Panels in this figure  are the same as in Fig.~\ref{fig:sclapp1}.
From top to bottom:
the superclusters SCl~474, SCl~512, SCl~525,
SCl~530, SCl~779, and SCl~827. 
}
\label{fig:sclapp3}
\end{figure}

In this section we present the distribution of 
galaxies and rich groups/clusters  in the sky  as well 
as the fourth Minkowski functionals and morphological signatures
for  superclusters with less than 950 member galaxies
(Fig.~\ref{fig:sclapp1}, ~\ref{fig:sclapp2}, and ~\ref{fig:sclapp3}). 
In all figures
the left panels show the distribution of galaxies in the sky (dots). 
Circles in these panels mark the location of groups with at least 
30 member galaxies. The size of a circle is proportional to the size of a group
in the sky. 
The middle panels show the fourth Minkowski functional $V_3$ vs. the mass fraction
$mf$, and the right panels show
the shapefinders $K_1$ (planarity) and $K_2$ (filamentarity) 
for the supercluster (the morphological signature).
Filled circles in the right panels mark the value of the mass fraction 
$mf=0.7$. In the right panels the mass fraction increases 
anti-clockwise along the curves.
As in the previous section, we do not
show the figures for superclusters, for which
the value of the fourth Minkowski functional $V_3 = 1$ 
over the whole mass fraction interval.

\end{appendix}

\end{document}